\documentclass[useAMS,usenatbib]{mn2e}
\usepackage{url}
\usepackage[colorlinks=true,citecolor=blue]{hyperref}
\usepackage{graphicx,graphics,color}
\usepackage{amssymb}    
\usepackage{array}
\usepackage[bottom]{footmisc}
\usepackage[pagewise]{lineno}
\usepackage[T1]{fontenc}
\usepackage{ae,aecompl}

\newcommand\chandra{{\it Chandra}}

\newcommand\xmm{{\it XMM-Newton}}

\newcommand\kms{\ifmmode {\rm~km\ s}^{-1} \else ~km s$^{-1}$\fi}
\newcommand\Hunit{\ifmmode {\rm~km\ s}^{-1}\ {\rm Mpc}^{-1}
        \else ~km s$^{-1}$ Mpc$^{-1}$\fi}
\newcommand\ctssec{\ifmmode {\rm~count\ s}^{-1} \else ~count s$^{-1}$\fi}
\newcommand\ergsec{\ifmmode {\rm~erg\ s}^{-1} \else
        ~erg s$^{-1}$\fi}
\newcommand\funit{\ifmmode {\rm~erg\ s}^{-1}\;{\rm cm}^{-2} \else
        ~ergs s$^{-1}$ cm$^{-2}$\fi}
\newcommand\phflux{\ifmmode {\rm~photon\ s}^{-1}\;{\rm cm}^{-2}
        \else   ~photon s$^{-1}$ cm$^{-2}$\fi}
\newcommand\efluxA{\ifmmode {\rm~erg\ s}^{-1}\;{\rm cm}^{-2}\;{\rm
        \AA}^{-1} \else ~erg s$^{-1}$ cm$^{-2}$ \AA$^{-1}$\fi}
\newcommand\efluxHz{\ifmmode {\rm~erg\ s}^{-1}\;{\rm cm}^{-2}\;{\rm
        Hz}^{-1} \else ~erg s$^{-1}$ cm$^{-2}$ Hz$^{-1}$\fi}
\newcommand\cc{\ifmmode {\rm~cm}^{-3} \else cm$^{-3}$\fi}
\newcommand\FWHM{\ifmmode {\rm~FWHM} \else ${\rm~FWHM}$\fi}
\newcommand\Zsun{\ifmmode Z_{\odot} \else $M_{\odot}$\fi}
\newcommand\Lsun{\ifmmode L_{\odot} \else $L_{\odot}$\fi}

\newcommand\hbeta{\ifmmode {\rm H}\beta \else H$\beta$\fi}
\newcommand\Kalpha{\ifmmode {\rm K}\alpha \else K$\alpha$\fi}
\newcommand\nh{\ifmmode N_{\rm H} \else N$_{\rm H}$\fi}

\newcommand{\lum}{erg\,s$^{-1}$}

\newcommand{\mnras}{MNRAS}

\newcommand{\Msun}{\ensuremath{\rm M_{\odot}}}
\newcommand{\mac}{\rm Abell~3017}

\title[Abell 3017]{ A detailed study of X-ray cavities in the intracluster environment of the
 cool core cluster Abell~3017} 
\author[Pandge et. al.]{M. B. Pandge$^{1}$\thanks{mbpandge@associates.iucaa.in, mbpandge@gmail.com}, 
Biny Sebastian$^{3}$, 
Ruchika Seth$^{2}$, 
Somak Raychaudhury$^{2,4}$\\
$^{1}$DST-INSPIRE Faculty, Dayanand Science College, Barshi Road, Latur 413512, Maharashtra, India\\
$^{2}$Inter University Centre for Astronomy and Astrophysics, Pune 411007, Maharashtra, India\\
$^{3}$National Centre for Radio Astrophysics (TIFR), Pune 411007, Maharashtra, India\\
$^{4}$School of Physics and Astronomy, University of Birmingham, Birmingham B15~2TT, UK\\
}
\begin{document}
\maketitle
\begin{abstract}
We present a detailed analysis of a cool-core galaxy cluster Abell~3017, at a redshift of $z\!=\! 0.219$, which has been identified to be merging with its companion cluster Abell~3016. This study has made use of X-ray ({\it Chandra}), UV (GALEX), optical (ESO/VLT), mid-infrared (WISE) and radio uGMRT observations of this cluster. Using various image processing techniques, such as unsharp masking, 2-d fits using Beta models, contour binning and the use of surface brightness profiles, we show the existence of a pair of X-ray cavities, at a projected distance of ~$\sim$20\arcsec (70\,kpc) and  ~$\sim$16\arcsec (57\,kpc), respectively from the core of Abell~3017. We also detect an excess of X-ray emission located at 25\arcsec $\sim$(88\,kpc) south of the centre of \mac, is likely due  to the bulk motions in the ICM either by gas sloshing or ram-pressure striping due to a merger. We find that the radio lobes are responsible for the observed X-ray cavities detected in this system. The lower values of mid-IR WISE colour [W1-W2] and [W2-W3] imply that the central BCG of Abell~3017 is a star-forming galaxy. The current star formation rate of the central BCG, estimated  from the ${\rm H\alpha}$ and {\it GALEX} FUV luminosities,  are  equal to be $\sim 5.06\pm 0.78$ \Msun yr$^{-1}$ and $\sim 9.20\pm 0.81$ \Msun yr$^{-1}$, respectively. We detect, for the first time, a radio phoenix $\sim$150\,kpc away from the radio core, with a spectral index of ($\alpha \!\leq\! -1.8$). We also report the detection of $\rm~Pa_\alpha$ emission in this cluster using ESO VLT {\tt SINFONI} imaging data. 
\end{abstract}


\section[1]{Introduction}
\begin{figure*}
\begin{center}
\includegraphics[width=16cm,height=8cm]{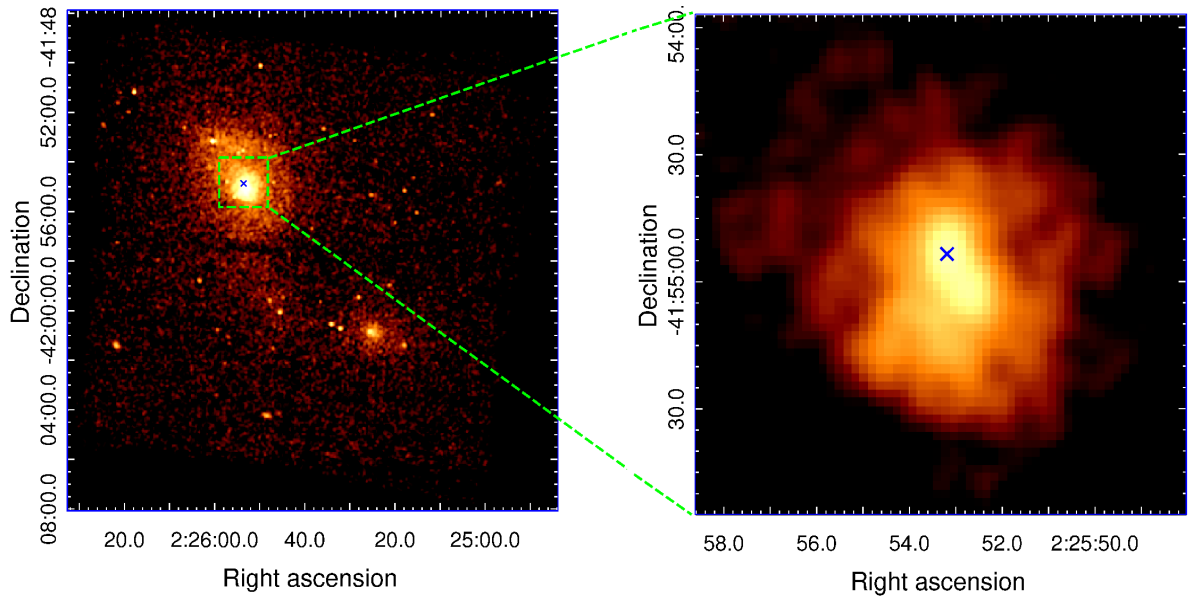}
\caption{ {\it Left panel}: A {\it Chandra} mosaic image of all the observations summarised in Table~\ref{tab1} of the galaxy cluster \mac, in the energy range 0.5$-$7\,keV. The image has been binned by a factor of 4 (4 pixel $\sim$ 2\arcsec). {\it Right panel}: The central 2\arcmin $\times$ 2\arcmin zoomed-in region of \mac. In each panel the X-ray peak position is indicated by a blue cross.}
\end{center}
\label{fig1}
\end{figure*}


X-ray observations of clusters of galaxies 
performed with  \chandra\ and \xmm\ have amply demonstrated that in spite of
the fact that the cooling times in their core are significantly shorter than the Hubble time, relatively little gas actually cools below about 1~keV \citep[e.g.][]{peterson2003}. Subsequent work on larger samples have demonstrated 
this is poorer clusters and groups as well \citep[e.g.][]{giacintucci2011,osullivan2011a}.
It is widely accepted now that 
radio sources residing in the central dominant elliptical galaxies
in groups and clusters (namely
active galactic nuclei, AGN) can reheat the inter-galactic gas via a variety of mechanisms \cite[][ and references therein]{mcnamara2007,fabian2012,2016NewAR..75....1S}.

The most common evidence of this interaction between the AGN and the intergalactic medium (IGM) comes from high-resolution X-ray images, in particular, from \chandra, often obtained at the superb spatial resolution $\le$0.5\arcsec, which show the presence of cavities, seen in the form of decrements in the X-ray surface brightness, 
often coincident with the lobes of extended radio emission
\citep[e.g.][]{mcnamara2007,david2011}. These cavities, inflated from  radio lobes, provide an uncomplicated method for estimating the power output of the jets. The mechanical energy required to displace the hot gas can be estimated from the volume of the cavity and ambient pressure of the IGM, the timescale  estimated from dynamical considerations. These cavities heat the surrounding gas as they rise buoyantly through the IGM, and several studies have shown that the energy required to create these cavities are enough to suppress the cooling in both groups and clusters
\citep{2004ApJ...607..800B,dunn2005,rafferty2006,osullivan2011b}.

 X-ray cavities have been detected in a majority of the cool core clusters, and a significant fraction of groups with X-ray emission, and this could be underestimated since these cavities are not easily detectable in projection against the diffuse emission from the hot ICM \citep{2004ApJ...607..800B,2010ApJ...712..883D}. As in the radio lobes, these X-ray cavities  are usually seen in pairs. The observed
 elliptical- shaped depressions can be detected, for example, by subtracting smoothed models of the X-ray surface brightness, at the level of less than $\sim$20\% or so   \citep{2010MNRAS.407..321O,2012MNRAS.421..808P,2013Ap&SS.345..183P}. Moreover, these  cavity systems come in a wide range of sizes, with diameters from less than a \,kpc, as in M87 \citep{2005ApJ...635..894F}, to more than 200\,kpc, as in systems such as the clusters MS0735.6+7421 and Hydra~A \citep{2005Natur.433...45M,2005ApJ...628..629N}. 

In this paper, we present the results obtained from the analysis of {\it Chandra} X-ray, uGMRT radio and ESO VLT (optical and near-IR) observations of the merging galaxy cluster \mac~\citep[z=0.2195,][]{1999ApJ...514..148D}. Recently, X-ray and radio analyses of this merging galaxy cluster have been published by \cite{2017MNRAS.470.3742P}. They have identified X-ray cavities in this system at $\sim$20\arcsec from the centre. In another study, \cite{2016MNRAS.460.1758H} pointed out that the bright central galaxy (hereafter BCG) of this system has a quiescent optical morphology, and that its velocity dispersion is 390$\pm$30 ${\rm kms^{-1}}$. Furthermore, they estimated the ${\rm H_{\alpha}}$ luminosity for the central BCG to be (110$\pm17)\times10^{40}$ \lum~. 

\cite{2019A&A...621A..77C} have published more interesting results using {\it Chandra} X-ray and ESO/MPG 2.2m telescope optical data and suggested two different evolutionary stages of the cluster merger based on the amount of gas present in the bridge region.  They have detected internal disturbances in the ICM of ~\mac , cold front at $\sim$ 150 kpc and shock front at $\sim$ 430kpc towards the south from its core.

This paper reports the results of a detailed analysis of new uGMRT radio observations, along with archival X-ray, optical and near-IR observations to investigate in detail the morphology of the core, the temperature structure of the IGM, and X-ray cavities, thereby enabling us to understand the nature of AGN feedback operating in in this merging cluster system.

The structure of this paper is as follows. In \S2. we describe the observational and data reduction strategy briefly for X-ray, radio, optical and near-IR data. X-ray spatial and spectral analysis and radio images and spectral analysis are described in \S3. and \S4. Results and conclusions of the study are outlined in \S5. and \S6., respectively. We assume $H_0$ = 70 km\, s$^{-1}$ Mpc$^{-1}$, $\Omega_M$=0.27 \& $\Omega_{\Lambda}$=0.73 translating to a scale of 3.554\,kpc\,arcsec$^{-1}$ at the redshift z=0.2195 of \mac. All spectral analysis errors are 90$\%$ confidence, while all other errors are 68 $\%$ confidence. Other global parameters of \mac has been tabulated in Table~\ref{basicpro}.

\begin{table}
\caption{The galaxy cluster \mac}
\begin{tabular}{@{}llrrrrlrlr@{}}
\hline
\hline
Other Name& PSZ1 G256.55-65.69\\
RA, Dec & 02:25:52, -41:54.5:00  \\
Redshift(z) & 0.2195 \\
Distance(Mpc) & 1045 \\
Plate scale& 3.55 \,kpc\, arcsec$^{-1}$\\
\hline
\hline
\end{tabular}
\label{basicpro}
\end{table}
\begin{table}
\begin{center}
  \caption{{\it Chandra} Observations of \mac.}
  \begin{tabular}{lcccrrlclc}
    \hline
    \hline
     &Observing &CCDs &Starting &Total& Clean \\
    ObsID &Mode &on &Date  &Time& Time \\
     & & &  &(ks)& (ks)\\
    \hline
    \hline
    15110 &VFAINT &0,1,2,3,6  & 2013-05-01 &15 & 14.27 \\
    17476 &VFAINT &0,1,2,3,6  & 2015-04-21 &14 & 13.74\\
    \hline
    \hline
    \end{tabular}    
  \label{tab1}
\end{center}
\end{table}

\begin{figure*}
\center
\includegraphics[width=160mm,height=140mm]{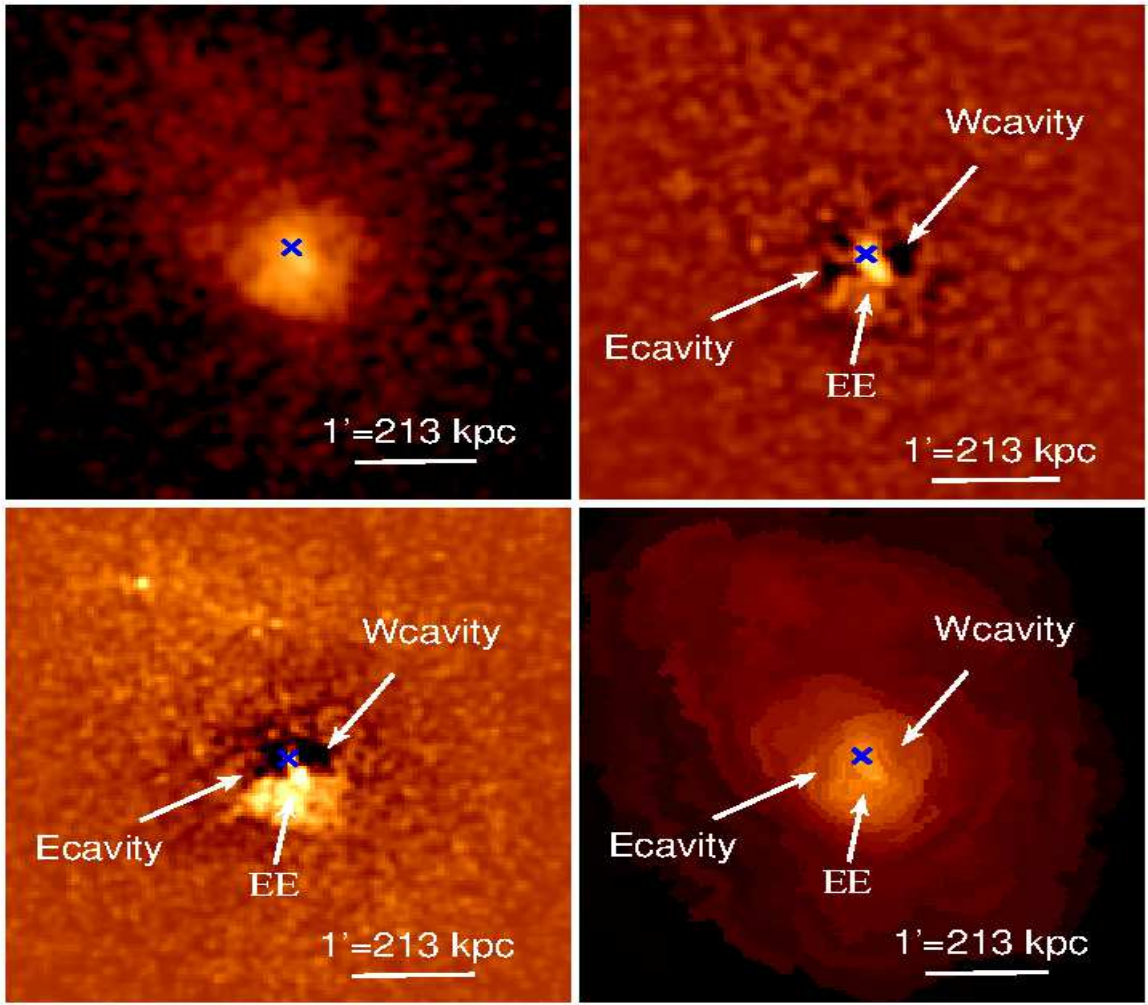}
\caption{{\it Top left panel:} Background-subtracted, exposure-corrected central 5\arcmin $\times$ 5\arcmin {\it Chandra} image of the galaxy cluster \mac, in the 0.7$-$2.0 keV energy band, smoothed by Gaussian width of $\sigma$=3\arcsec. {\it Top right panel:} Unsharp-masked image of \mac, after subtracting a 1$\sigma$ Gaussian kernel-smoothed image from that smoothed with a 5$\sigma$ Gaussian kernel. {\it Bottom left panel:} Elliptical 2-d beta model-subtracted residual image, exhibiting an excess X-ray emission towards the south of the core of the cluster.  {\it Bottom right panel:} The spatially binned image produced using the {\tt CONTBIN} algorithm with S/N threshold of 5. The image is smoothed by 20 (S/N). In all panels the blue cross mark indicates the position of the X-ray peak, while arrows show the positions of the detected X-ray cavities (Ecavity and Wcavity), and the excess X-ray emission region (EE) in \mac.}
\label{fig2}
\end{figure*} 

\begin{figure}
\includegraphics[width=80mm,height=75mm]{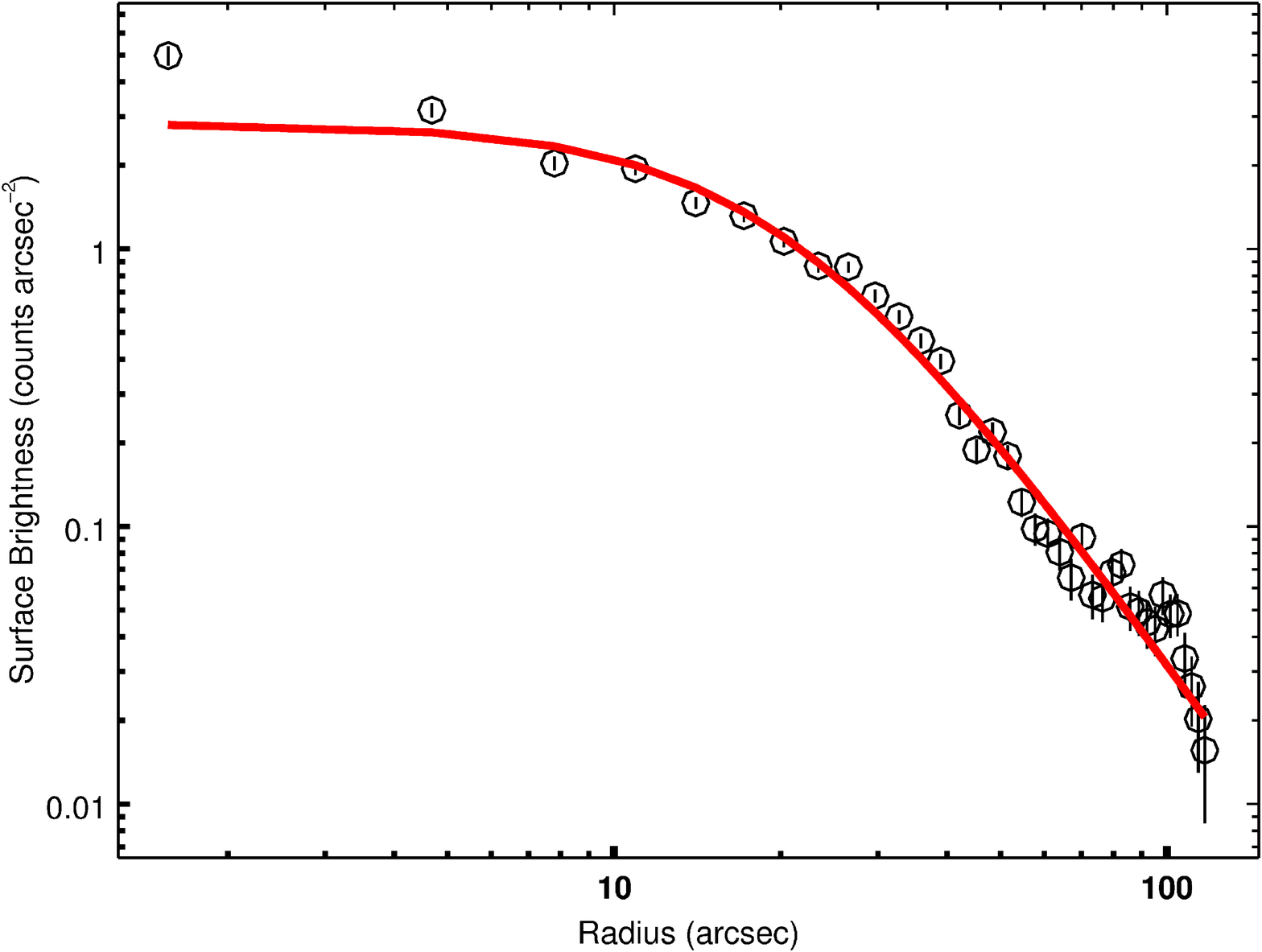}
\caption{ An azimuthally-averaged, background-subtracted surface brightness profile of the galaxy cluster \mac~in the energy range 0.3$-$7.0\,keV. The continuous red line represents the best-fit $\beta$ model for the azimuthally averaged surface brightness profile.} 
\label{fig3}
\end{figure}

\begin{figure*}
\includegraphics[width=80mm,height=75mm]{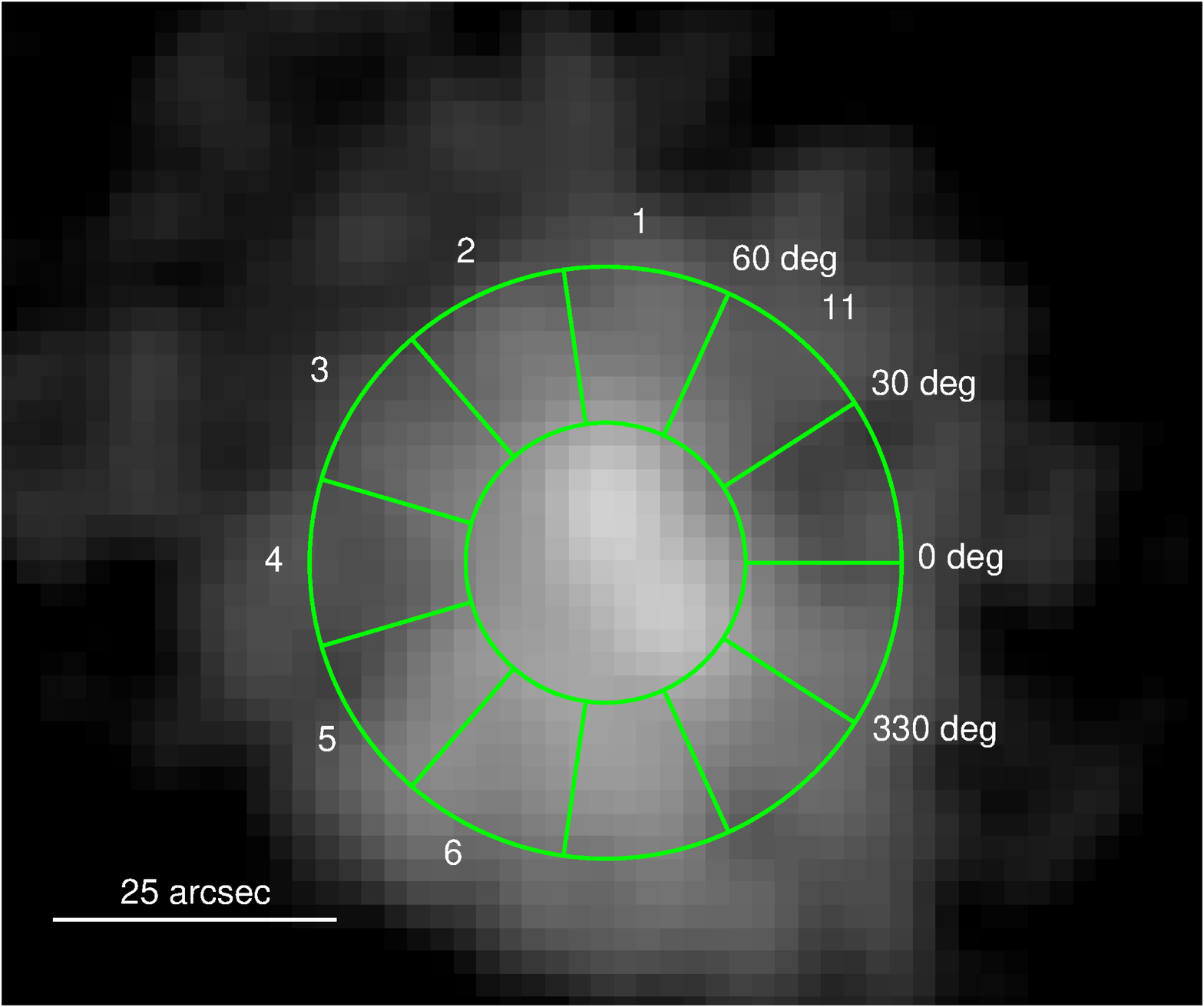}
\includegraphics[width=80mm,height=75mm]{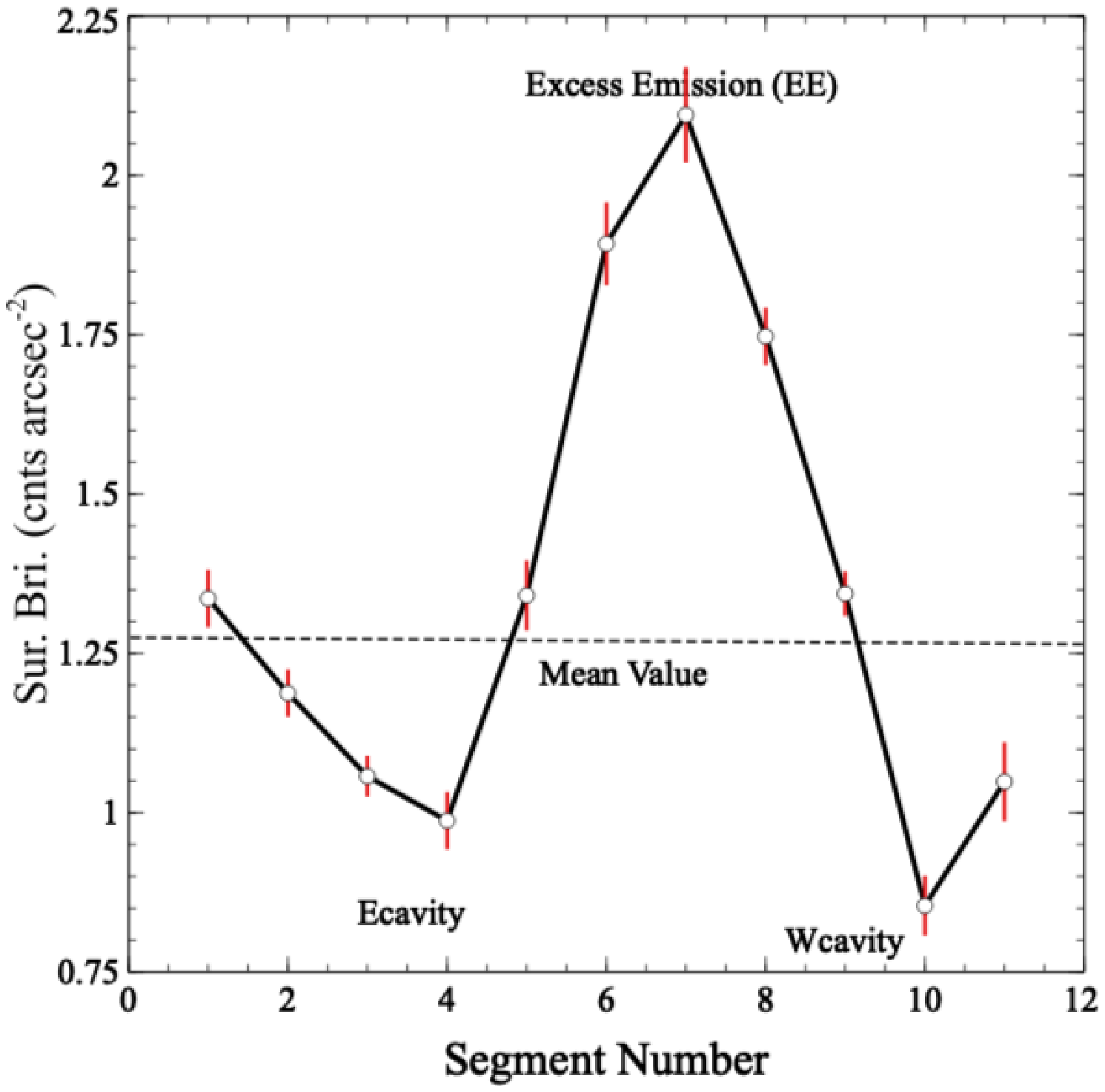}
\caption{{\it Left panel:} {\it Chandra} image of \mac~in the energy range of 0.5$-$7.0\,keV band, on which 11 segments overlaid. These segments were used to extract the surface brightnesses. {\it Right panel:} Shows a plot of X-ray surface brightness versus segment numbers. The X-ray deficiencies along the segments 4, 5 and 10 depict X-ray cavities, while segments 6, 7 and 8 represents an excess emission (EE). The horizontal dashed line indicate the mean value of X-ray photons.}
\label{fig4}
\end{figure*}

\section{Observations and Data Reduction}
\subsection{X-ray observations}
\label{The X-ray}
\mac~has been observed twice by \chandra\ X-ray Observatory, between  May 2013 and April 2015, for an effective exposure time of 29\,ks (ObsID 15110 and 17476, Table~\ref{tab1}). We reprocessed each observation using the \texttt{{CHANDRA\_REPRO}} task available within the CIAO~4.8 package. We used the calibration unit CALDB 4.7.2 provided by the {\it Chandra} X-ray Center (CXC), following the standard \chandra~ data-reduction threads {\tt http://cxc.harvard.edu/ciao/threads/index.html}. The background was modelled according to the procedure outlined in \citet{2017MNRAS.472.2042P}, and the resultant file was used to construct the background-subtracted images and for the spectral fitting procedure. The log-scale \chandra\ mosaic image of \mac\ in the energy range 0.5$-$7.0\,keV, constructed from all the ObsIDs summarized in Table~\ref{tab1} is shown in Fig.~\ref{fig1} (left panel). To visualize the central part of \mac,~we zoom in to the the inner 2\arcmin $\times$ 2\arcmin  region of the cluster in the same Figure (right panel). The peak X-ray position is marked by a blue cross symbol.

\subsection{uGMRT Radio Data}
Low-frequency radio observations of Abell\,3017 (32$\_$091) was carried out using the upgraded Giant Metrewave radio telescope (uGMRT) at Band-3 (250-500 MHz) and Band-5 (1200-1400 MHz). The details of the observations are given in Table~\ref{Obs}.
\begin{table}
 \begin{center}
 \caption{uGMRT Observation Details }
 \label{Obs} 
 \begin{tabular}{cccccl}
\hline
Central  & Flux  & Phase & Band- & Time on& Obs. \\ 
Frequency &\multicolumn{2}{c}{Calibrator} &width & source& Date\\ 
(MHz)&&&(MHz)&(h)&\\
\hline
400  &3C48 &0155$-$408 &200 & 3.0 & 16-Sep-2017\\
1350 &3C48 &0155$-$408 &200 & 1.7 & 18-Jun-2017\\
\hline
\end{tabular}
\end{center}
\end{table}
\subsubsection{uGMRT data analysis and imaging} 
\label{dataan}
Basic editing and calibration of the uGMRT data were done in the Astronomical Image Processing System (AIPS). Data corrupted by radio-frequency interference (RFI) and other instrumental issues were flagged out using standard AIPS procedures. After a few initial rounds of flux and phase calibration, the bandpass calibration was performed using the flux calibrator. Few channels in the multi-source file were averaged using task `SPLAT' to increase the signal to noise ratio and to reduce the data size. After the data were averaged, the already existing calibration tables were deleted, and the calibration was redone. The target source was then split out after applying the calibrations, and imaged using only a fraction of the whole bandwidth such that the fractional bandwidth being imaged was about 10\%. 
Three rounds of phase-only self-calibration were carried out. The gain solutions were applied to the entire band. The final calibrated file was then exported to CASA and a final image using the whole bandwidth was carried out using MS-MFS algorithm. $W$-projection was used to take care of $w$-term errors.

To obtain spectral information the raw uv-data were split into four sub-bands at Band-5 and six sub-bands at Band-3. Each of these subsets was also calibrated and imaged following standard AIPS routines. In our observations, The last three sub-bands at Band-3 were completely corrupted by RFI. Hence a total of seven images could be generated, to be used for a detailed spectral analysis across the band.

Since the uv-coverage and the resolution of Band-5 and Band-3 are very different, two in-band spectral index images were separately created from both the frequency bands. Yet another round of imaging with matched (u,v)-ranges to generate similar resolution images at various frequencies were carried out. The lower (u,v)-ranges of lower frequency sub-bands were matched to that of the highest frequency sub-band data in both Band-5 and Band-3. The higher frequency sub-band data sets were then tapered such that the resolution turned out to be similar. The matching of (u,v) ranges was done to make sure that the scales of emission that is being picked up is similar in all the sub-bands. 

We convolved the final sub-band images using a common circular beam in both the bands using task `CONVL' in AIPS. Since we had four sub-band points in Band-5, we performed a pixel-to-pixel fitting of the flux density as a function of frequency to generate the spectral index image. We used only the last two sub-band images at Band-3 to make the two-point spectral index image using the task `COMB' in AIPS. 

\subsubsection{GMRT Radio Data}
We have also used GMRT 235 and 610~MHz radio counter to show that the diffuse radio emission fills the X-ray cavity regions. More details about GMRT data analysis can be found in \citep{2017MNRAS.470.3742P}


\begin{table}
    \caption{Radio Imaging Parameters}
    \label{Image Parameters}
    \begin{tabular}{lccr}
        \hline
        Frequency & Resolution & Rms & PA \\
         (MHz) & & 8\% (mJy) & \\
        \hline    
        1426 & $4.43\times 1.94$ & 0.06 & 22.88 \\
        1375 & $4.75\times 2.12$ & 0.05 & 22.56\\
        1326 & $4.86\times 2.23$ & 0.03 & 24.16\\
        1275 & $5.27\times 2.48$ & 0.12 & 25.80\\
         384 &$13.11\times 6.95$ & 0.30 &  7.13\\
         349 &$16.01\times 8.86$ & 0.70 & 13.22\\
         317 &$18.3\times 10.31$ & 1.16 & 15.23\\
        \hline
    \end{tabular}
    \end{table}
    
\subsection{Optical and near-IR Observations}
\subsubsection{Optical Imaging}
We searched for optical data for \mac~in various archives and found  $\sim$300~s, optical R$-$Bess filter images obtained using the FORS1 instrument in the ESO-VLT archives [prog\_id=70.B-0440(A)]. The raw data for \mac~was obtained from the ESO archive and reduced using the standard dedicated pipeline `ESO Reflex'\footnote{\color{blue}https://www.eso.org/sci/software/reflex/}. The final science image is used to study the distribution of member galaxies around the BCG of \mac~ in \S~\ref{Sec: XOM}.
\subsubsection{Near-IR Imaging}
\label{IRI}
We also used the data from seeing-limited   K band  (FOV: 8\arcsec $\times$ 8\arcsec) observations from the ESO archives using the {\tt SINFONI} instrument (Proposal Nos. 086.A-0810(A)), with a total on-source exposure time of 600s. The standard pipeline which is provided by ESO was used for the data reduction. The red-shifted P$\alpha$ continuum image that was extracted from the {\tt SINFONI} data cube is shown in Fig.~\ref{mor} (Bottom right panel).

\section{Analysis of X-ray observations}

\subsection{X-ray Spatial Analysis}

To look for cavities, we use the background-subtracted 0.7$-$2.0~KeV X-ray images, produced as outlined in \S{\ref{The X-ray}}, shown in the top right panel of Fig.~\ref{fig2}.

We carefully searched for surface brightness decrements, hidden features and substructures in this  image, and could detect signs of X-ray surface brightness decrements in the inner 2\arcmin $\times$ 2\arcmin~ region (Fig.~\ref{fig1} right panel).  We used four distinct image processing techniques, namely unsharp masking, 1-D surface brightness profiles, 2-d $\beta$ model-fitting \citep[see][]{2010ApJ...712..883D,2012MNRAS.421..808P} and contour-binning \citep[see][]{2006MNRAS.371..829S}, to detect and explore the presence of substructures in the \chandra\ X-ray image. 

The unsharp-masked image of \mac, produced after subtracting a 1$\sigma$ Gaussian kernel-smoothed image from that smoothed with a wider 5$\sigma$ Gaussian kernel image \citep{2012MNRAS.421..808P}, is shown in Fig.~\ref{fig2} (Top right panel). This clearly shows the presence of two cavities as decrements on either side of the nucleus. A linear scale corresponding to 1\arcmin~ is superposed on all panels in this image. 
\begin{figure}
\centering
\includegraphics[width=80mm,height=80mm]{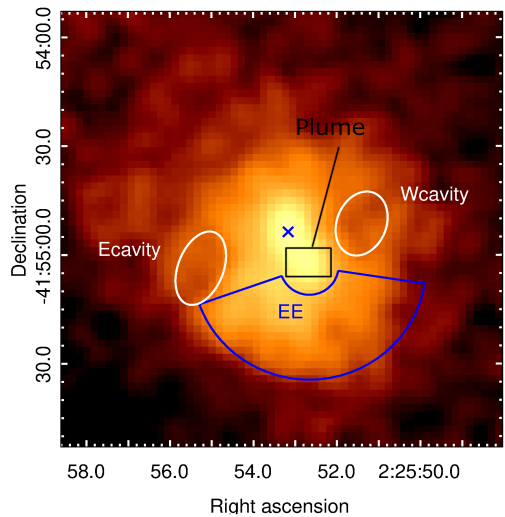}
\caption{The{\it Chandra} image of \mac~in the energy range of 0.5$-$7.0\,keV band (exposure-corrected, background-subtracted, point source removed, 2$\times$2\arcmin, smoothed by a Gaussian width $\sigma$=1.5\,\arcsec), on which the regions used for spectral extraction are overlaid (except the black box). In the same figure, the plume-like region is shown by a black box near the centre of the cluster. We comment on the excess emission seen in the region marked EE in \S5.3}
\label{fig5}
\end{figure}
\begin{figure}
\center
   \includegraphics[width=80mm,height=80mm]{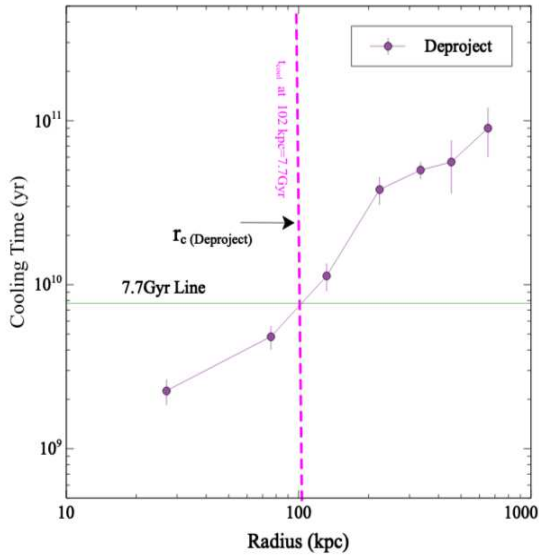}
   \caption{Deprojected cooling time profiles for \mac. The horizontal
solid line corresponds to the cooling time of 7.7 Gyr. The vertical dashed lines (magenta) represents  deprojected cooling radius of \mac~cluster at 7.7 Gyr.}
\label{CT}
\end{figure}

\begin{figure*}
\hspace{-1cm}
\includegraphics[width=60mm]{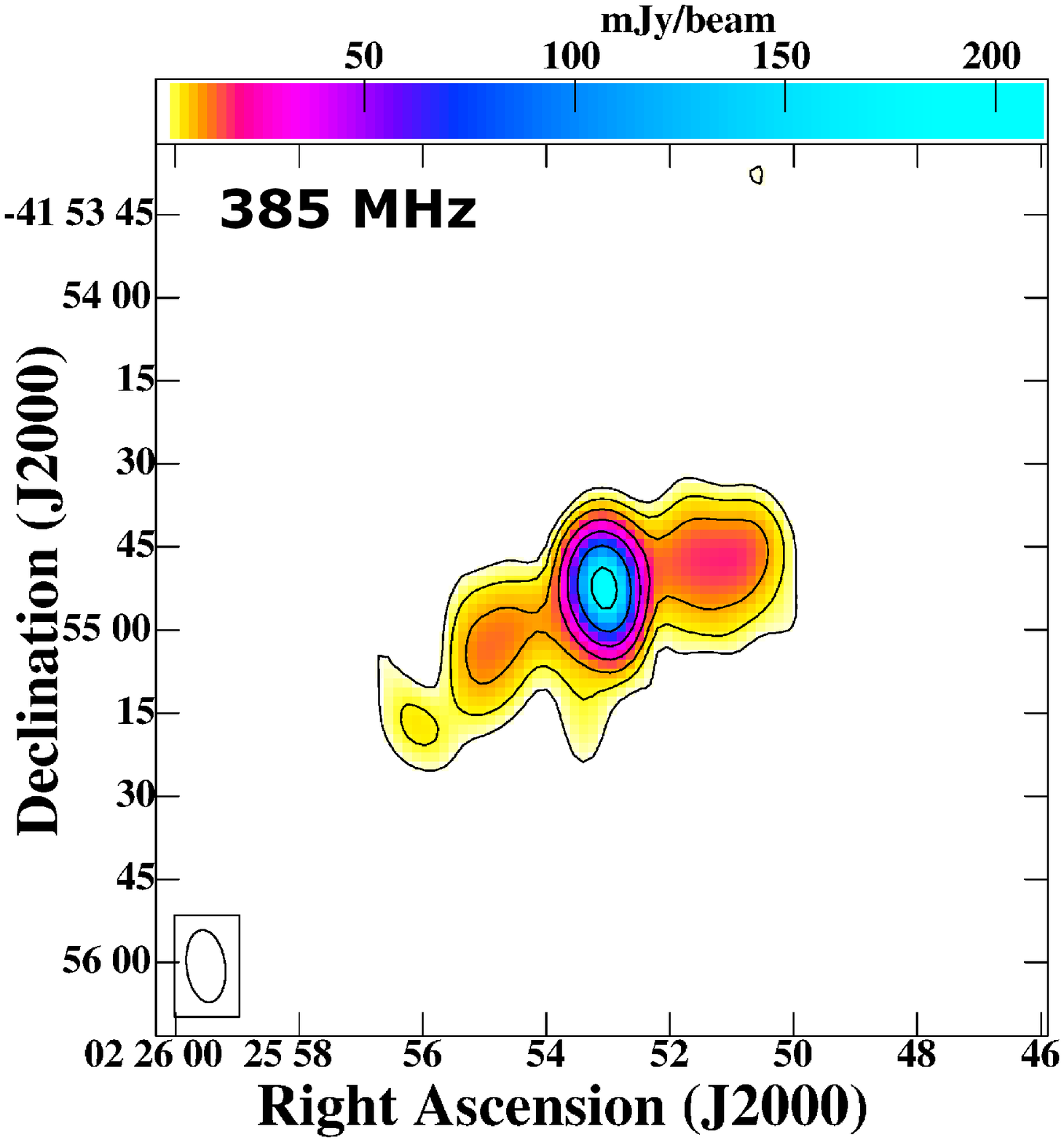}
\includegraphics[width=60mm]{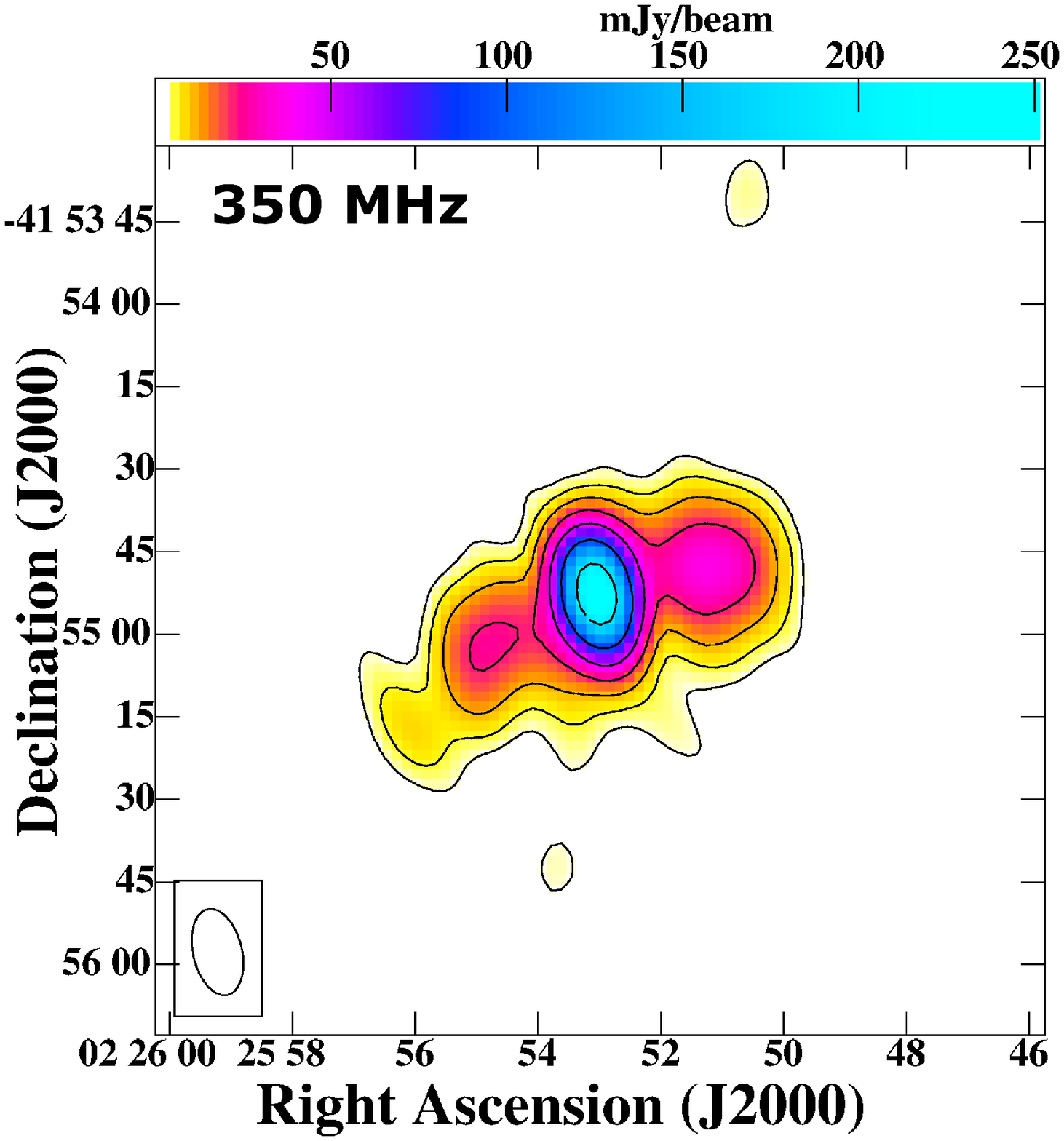}
\includegraphics[width=60mm]{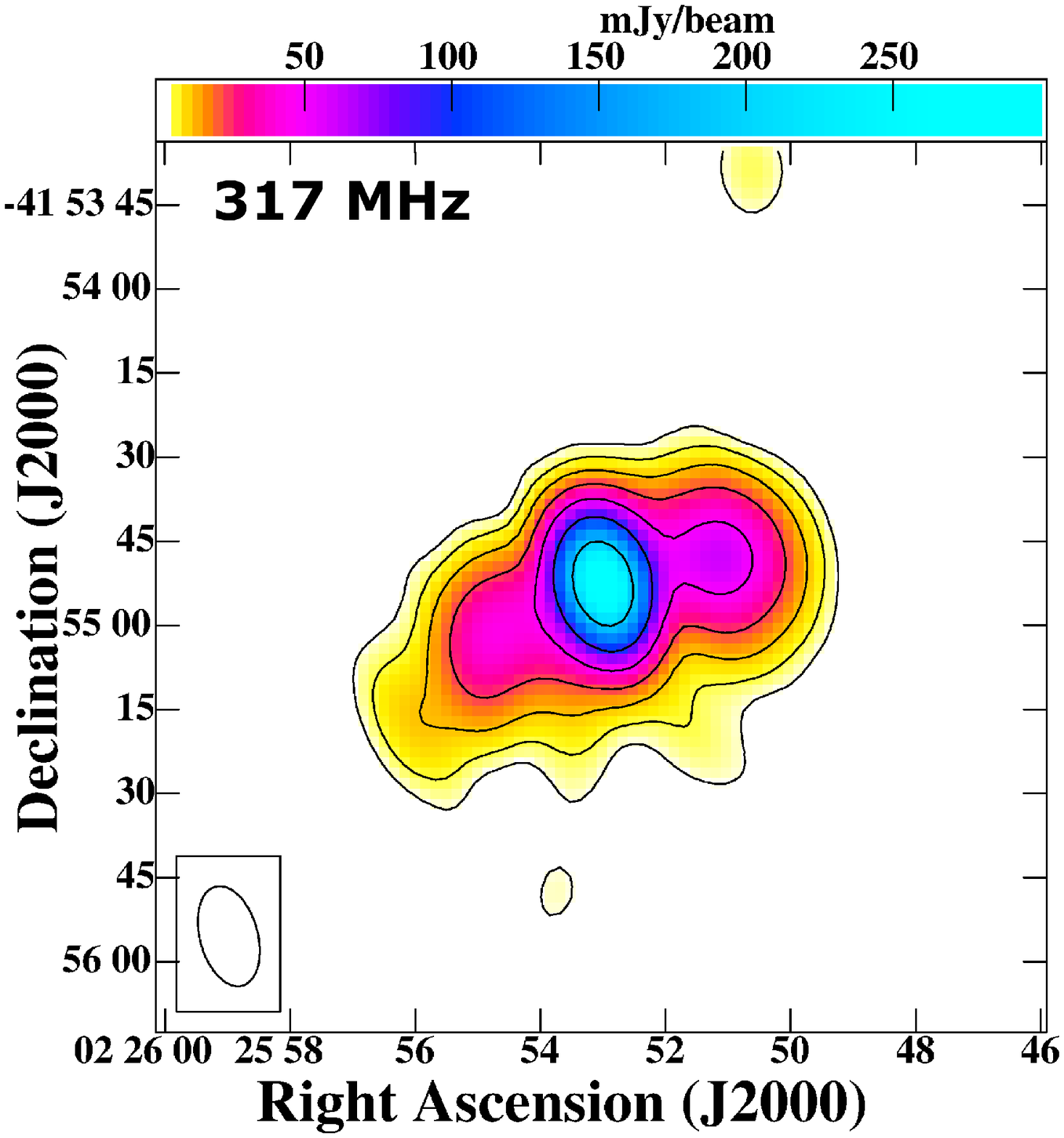}
\caption{Sub-band continuum radio images of Abell~3017 in uGMRT Band-3 (300-500 MHz). Left panel: Contour image at 385 MHz. The contour levels are 2.63 mJy $\times$ (-2, -1, 1, 2, 4, 8, 16, 32, 64). Middle panel: Contour image at 350 MHz. The contour levels are 2.85 mJy $\times$ (-2, -1, 1, 2, 4, 8, 16, 32, 64). Right panel: Contour image at 317 MHz. The contour levels are 2.93 mJy $\times$ (-2, -1, 1, 2, 4, 8, 16, 32, 64). }
\label{fig:pband}
\end{figure*}
\begin{figure*}
\includegraphics[width=65mm,height=65mm]{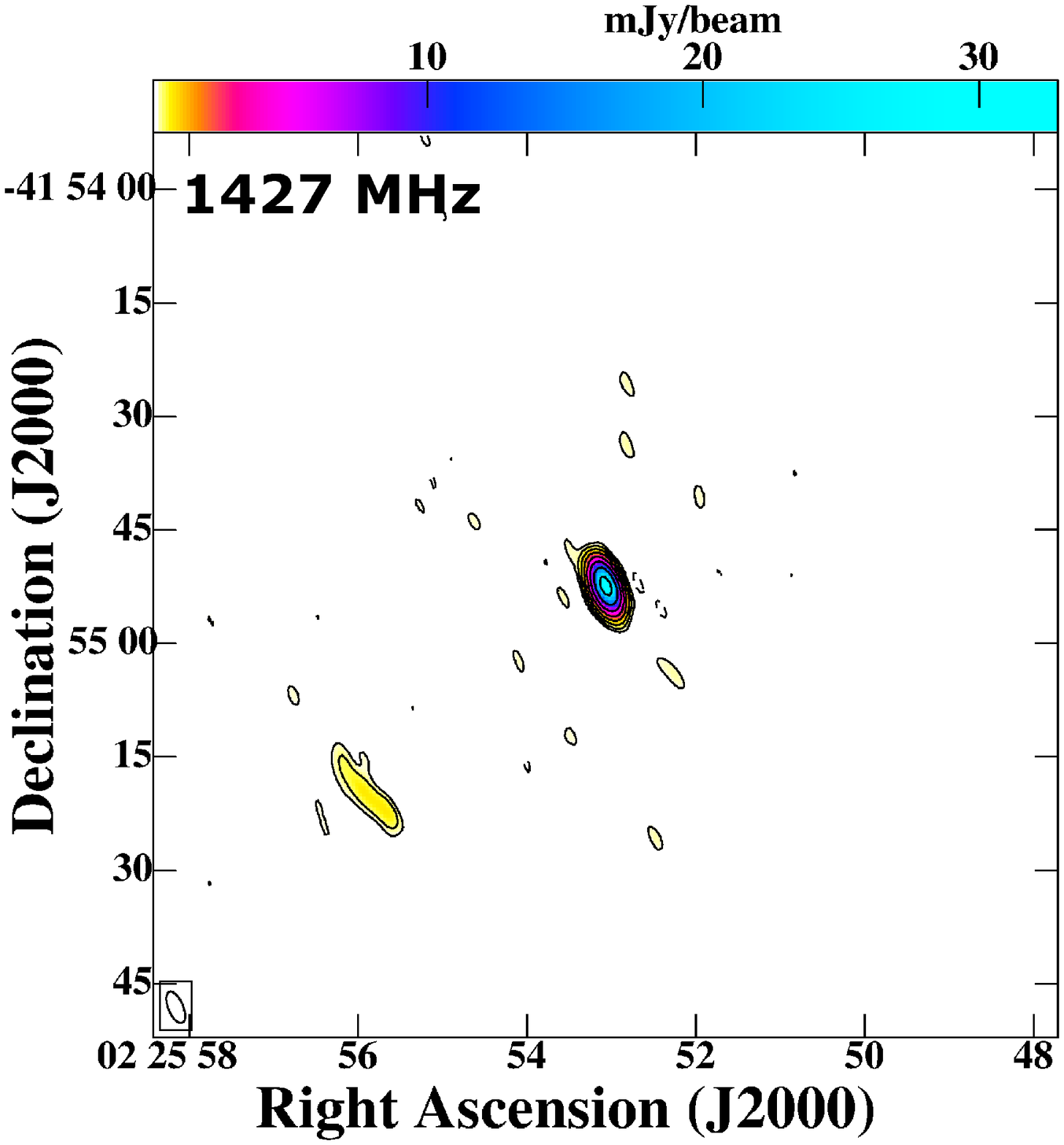}
\includegraphics[width=65mm,height=65mm]{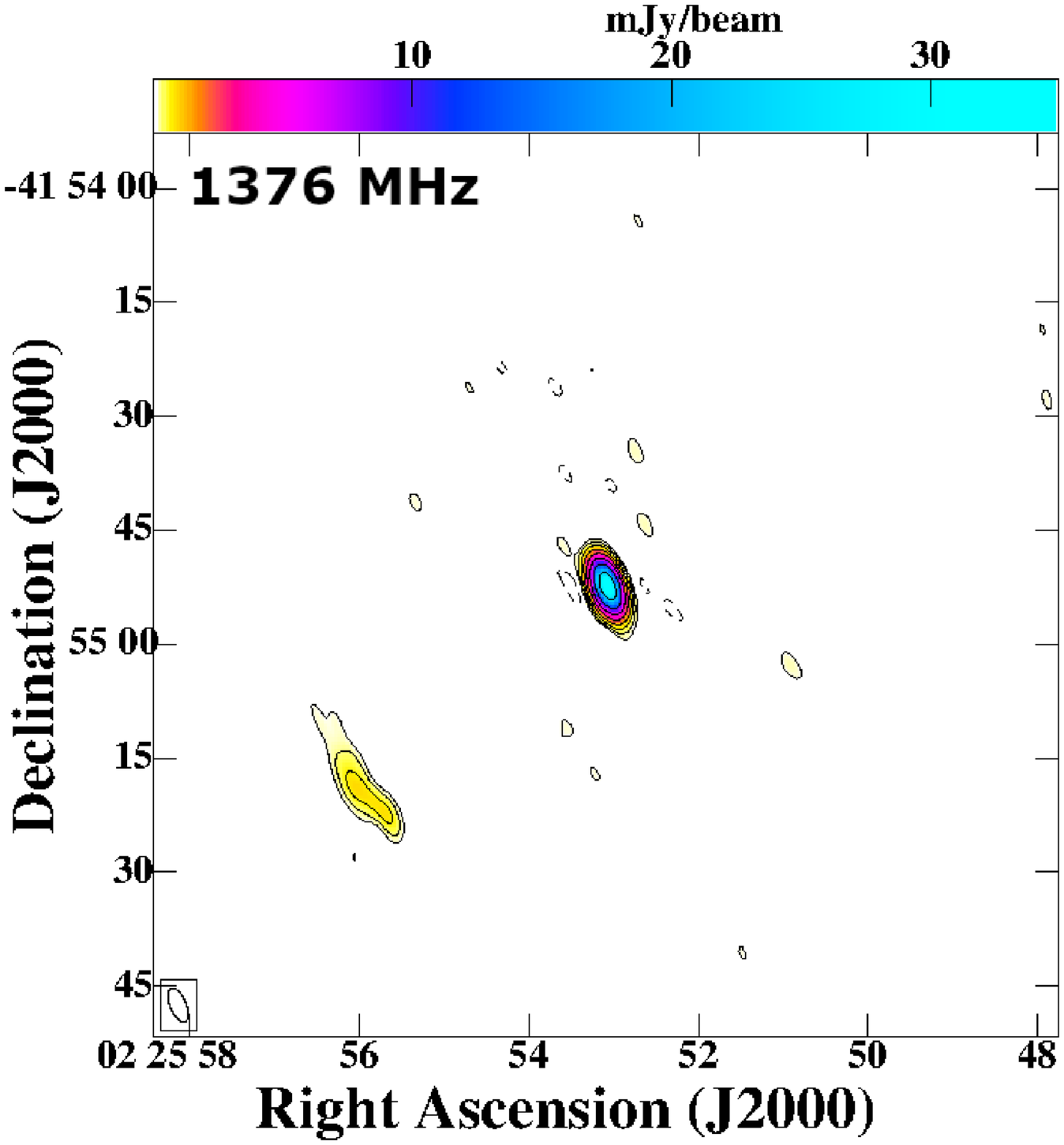}
\includegraphics[width=65mm,height=65mm]{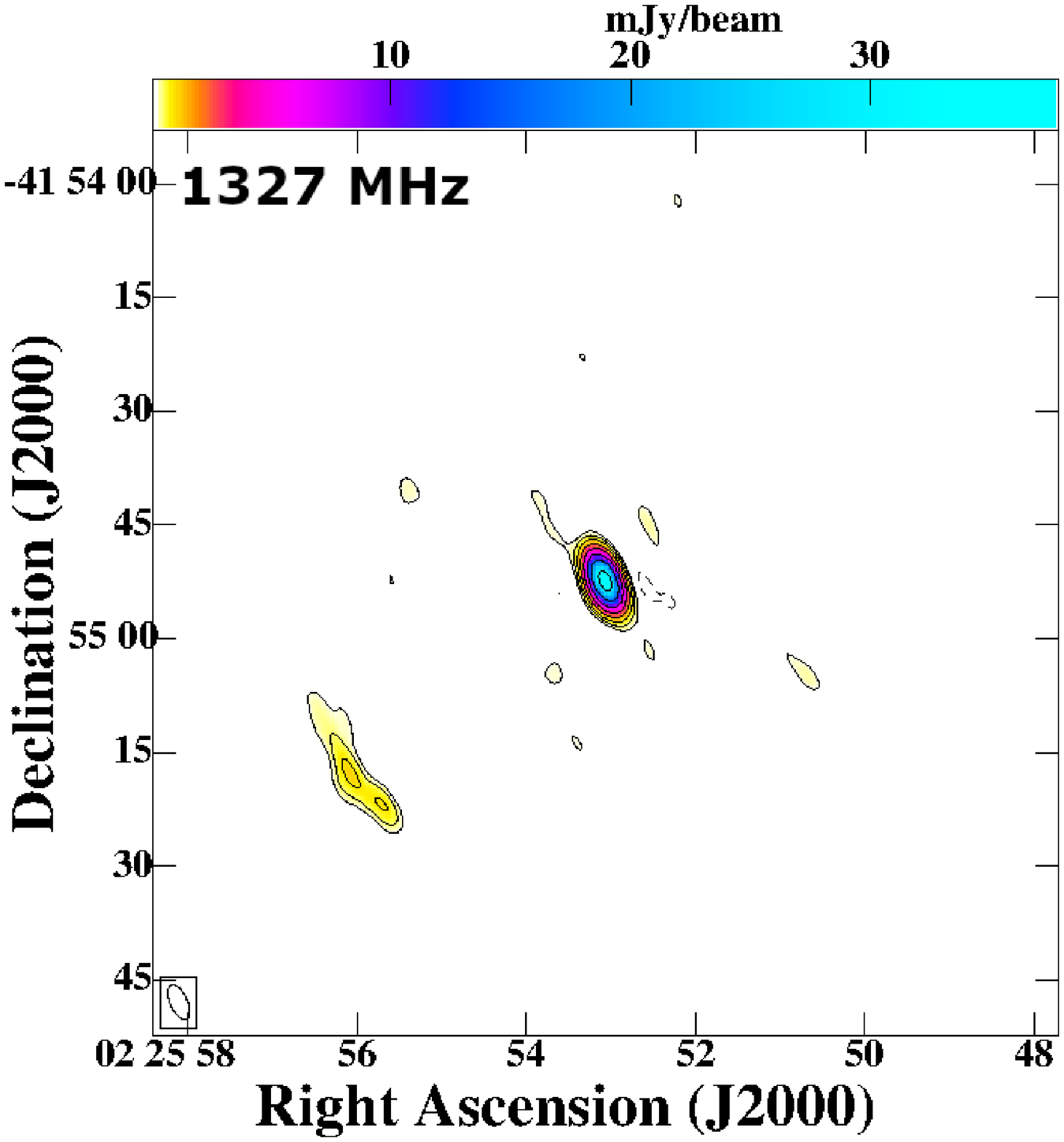}
\includegraphics[width=65mm,height=65mm]{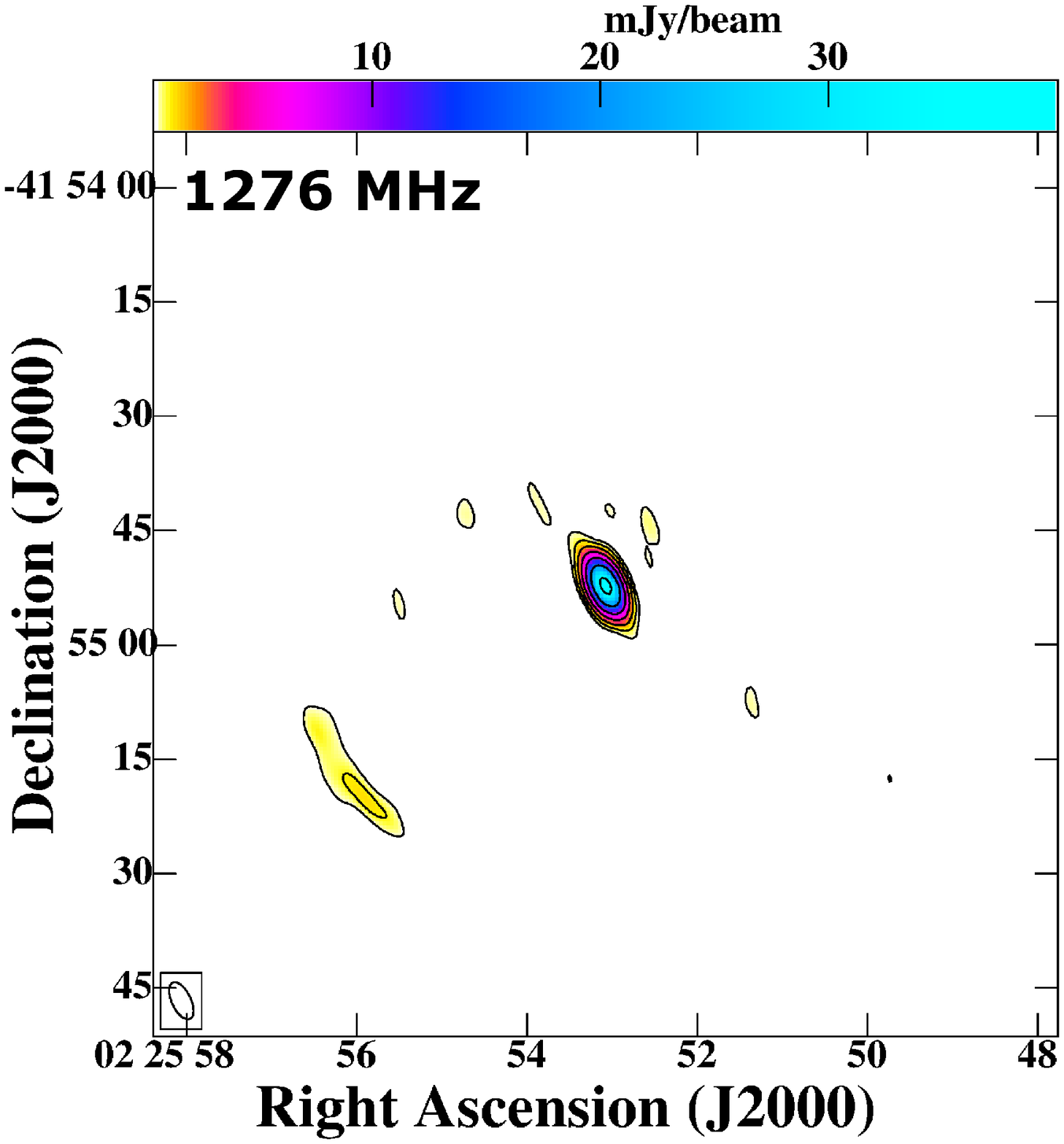}
\caption{Sub-band continuum radio images of Abell~3017 in uGMRT Band-5 (1250-1450 MHz). Top left panel: Contour image at 1427 MHz. The contour levels are 0.200 mJy $\times$ (-2, -1, 1, 2, 4, 8, 16, 32, 64). Top right panel: Contour image at 1376 MHz. The contour levels are 0.18 mJy $\times$ (-2, -1, 1, 2, 4, 8, 16, 32, 64). Bottom left panel: Contour image at 1327 MHz. The contour levels are 0.24 mJy $\times$ (-2, -1, 1, 2, 4, 8, 16, 32, 64). Bottom right panel: Contour image at 1276 MHz. The contour levels are 0.55 mJy $\times$ (-2, -1, 1, 2, 4, 8, 16, 32, 64). }
\label{fig:lband}
\end{figure*}


To confirm the reality of these cavities, we employ a different algorithm on the original image. Subtracting an elliptical 2-D beta model  \citep[][]{2010ApJ...712..883D,2012MNRAS.421..808P}, we show the resulting residual map in Fig.~\ref{fig2} (Bottom left panel).  In addition to the two decrements seen in the unsharp-masked image, we see in addition a bright knot to the south of the nucleus. The X-ray cavities are located at a projected distance of  $\sim$ 20\arcsec (70\,kpc) east (hereafter Ecavity) and $\sim$ 16\arcsec (57\,kpc) north (hereafter Wcavity) from the centre of \mac. 

Further, we used the contour-binning algorithm of \cite{2006MNRAS.371..829S} to construct a binned image, after removing point sources over a S/N threshold of 5, from the 0.3$-$5.0\,keV X-ray image, smoothing the image by 20 (S/N). The output of this exercise is shown in Fig.~\ref{fig2} (Bottom right panel). The X-ray brightness decrements (hereafter X-ray cavities) as well as an excess X-ray emission (hereafter ``EE") seen $\sim$ 25\arcsec (88\,kpc) to the south from the centre (highlighted by white arrows) are also evident. 

\subsubsection{Surface Brightness Profile}
\label{SBD}
We investigated the significance of the detected X-ray cavities by extracting the projected azimuthally-averaged surface brightness profile, by constructing 38 circular annuli centred on the adopted core of \mac.  This surface brightness profile is shown in Fig.~\ref{fig3}. In this plot, the extracted azimuthally averaged data points are indicated by open black circles and the best-fit model by a solid red line. The profile was then fitted with a 1D $\beta$ model \citep{1962AJ.....67..471K}, the best-fitting parameters values being $\beta=0.63\pm0.017$ and $R_{c}=71\pm$2\,kpc.  The best-fit $\beta$ value is consistent with the typical value found for clusters \citep[$\sim$0.64, e.g.][]{1999ApJ...517..627M}.  Further, to confirm the deficiencies in the X-ray emission along the cavity regions, we carried out the  X-ray count statistics by extracting the counts form the 11 segments within annular regions of 12-25\arcsec, as shown in Fig.~\ref{fig4}, left panel. Moreover, the right panel of same figure shows the variation of surface brightnesses versus the number of segments.  We find the considerable amount of decrements in the X-ray counts relative to the mean count value (calculated using the segments 1, 2, 5 and 9). From the same figure it is also evident that counts from the segments 6, 7 and 8 are 
higher than that of the mean value, probably because of an excess emission seen along that direction. 


\begin{figure}
    \centering
   \hspace*{-0.5cm}
    \includegraphics[width=85mm]{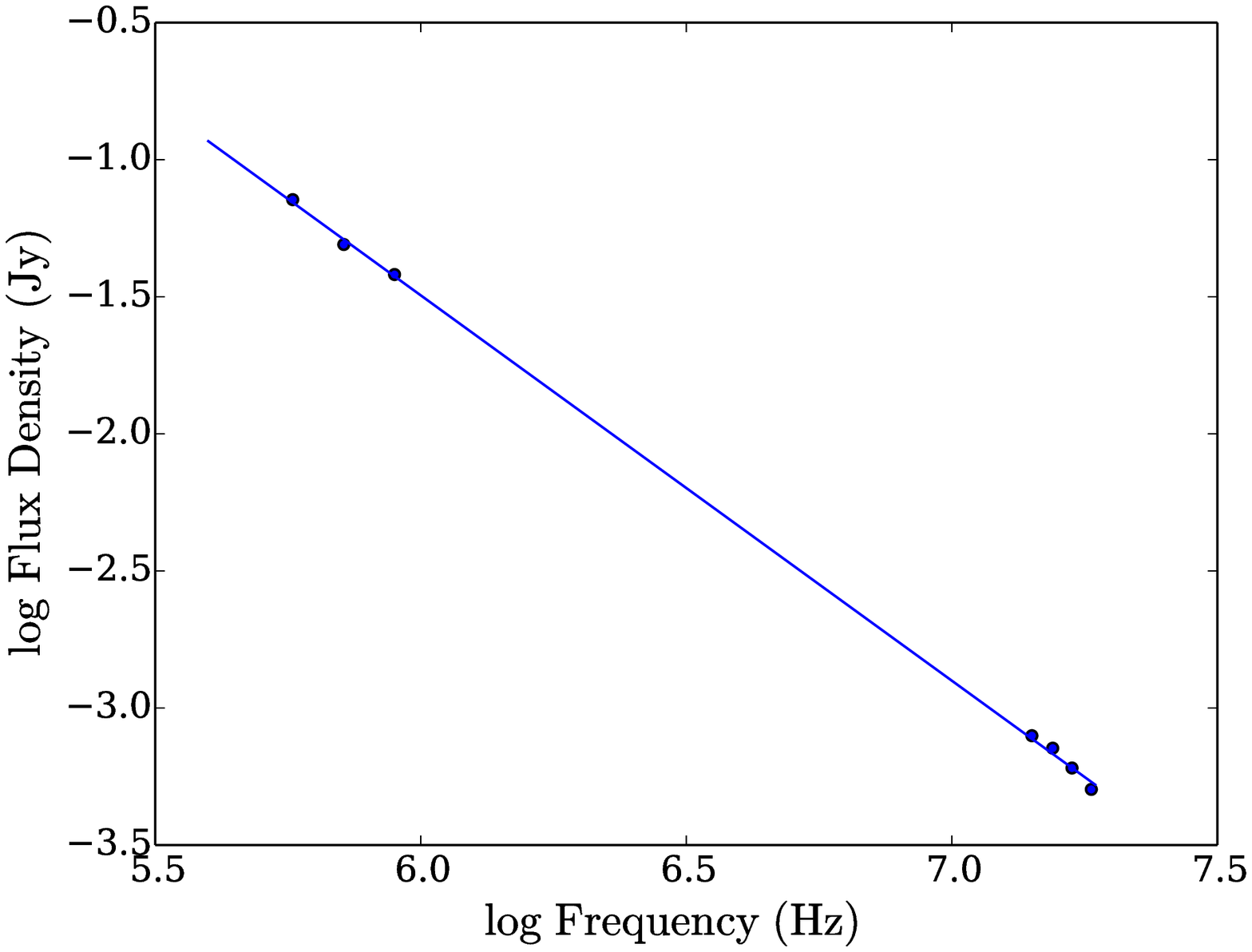}
   \caption{Radio spectrum of the core of Abell 3017.}
    \label{fig:spectral_index_bcg}
\end{figure}
\begin{figure*}
\vspace*{1cm}
    \includegraphics[height=80mm,width=80mm]{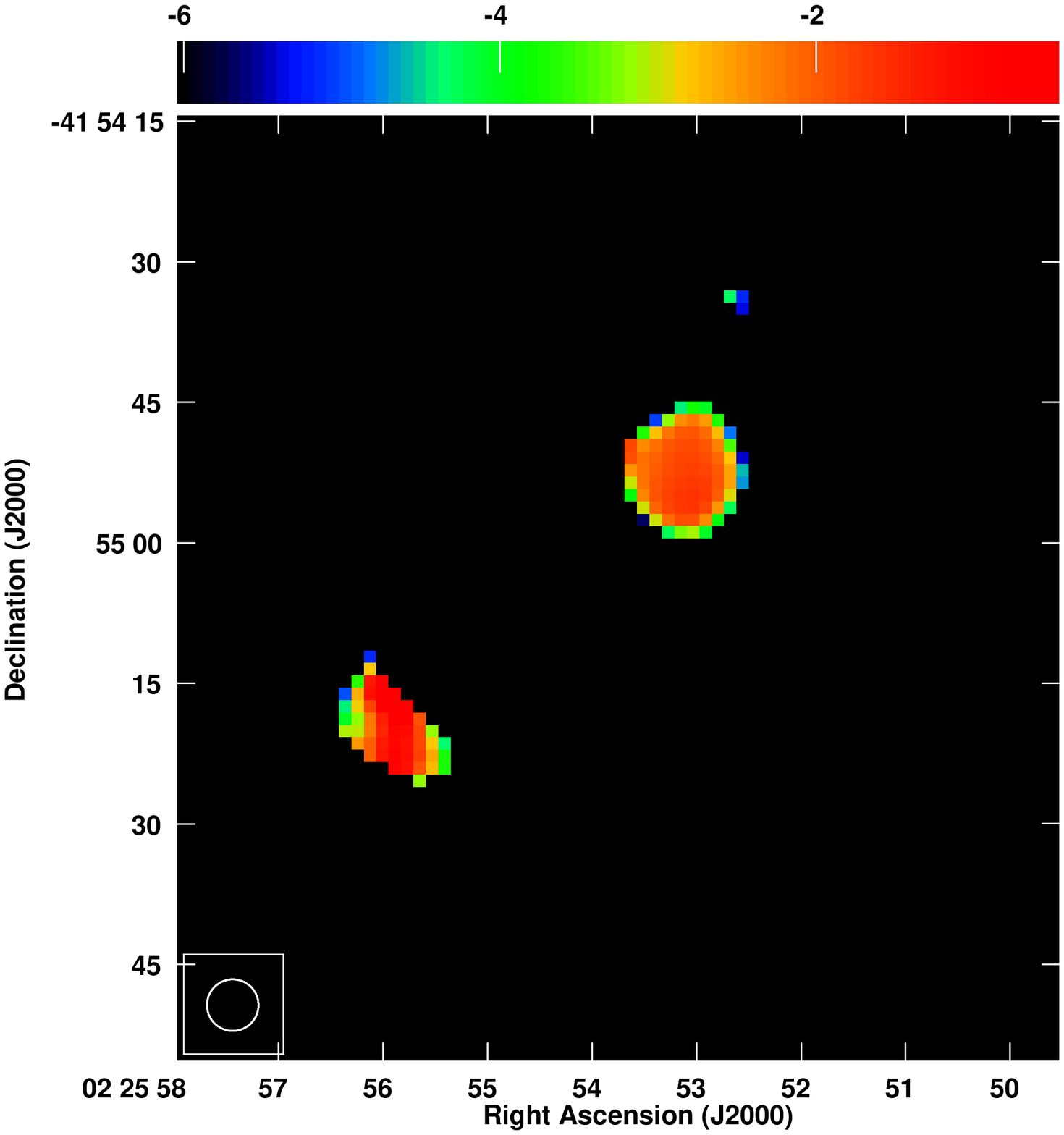}
   \includegraphics[height=80mm,width=80mm]{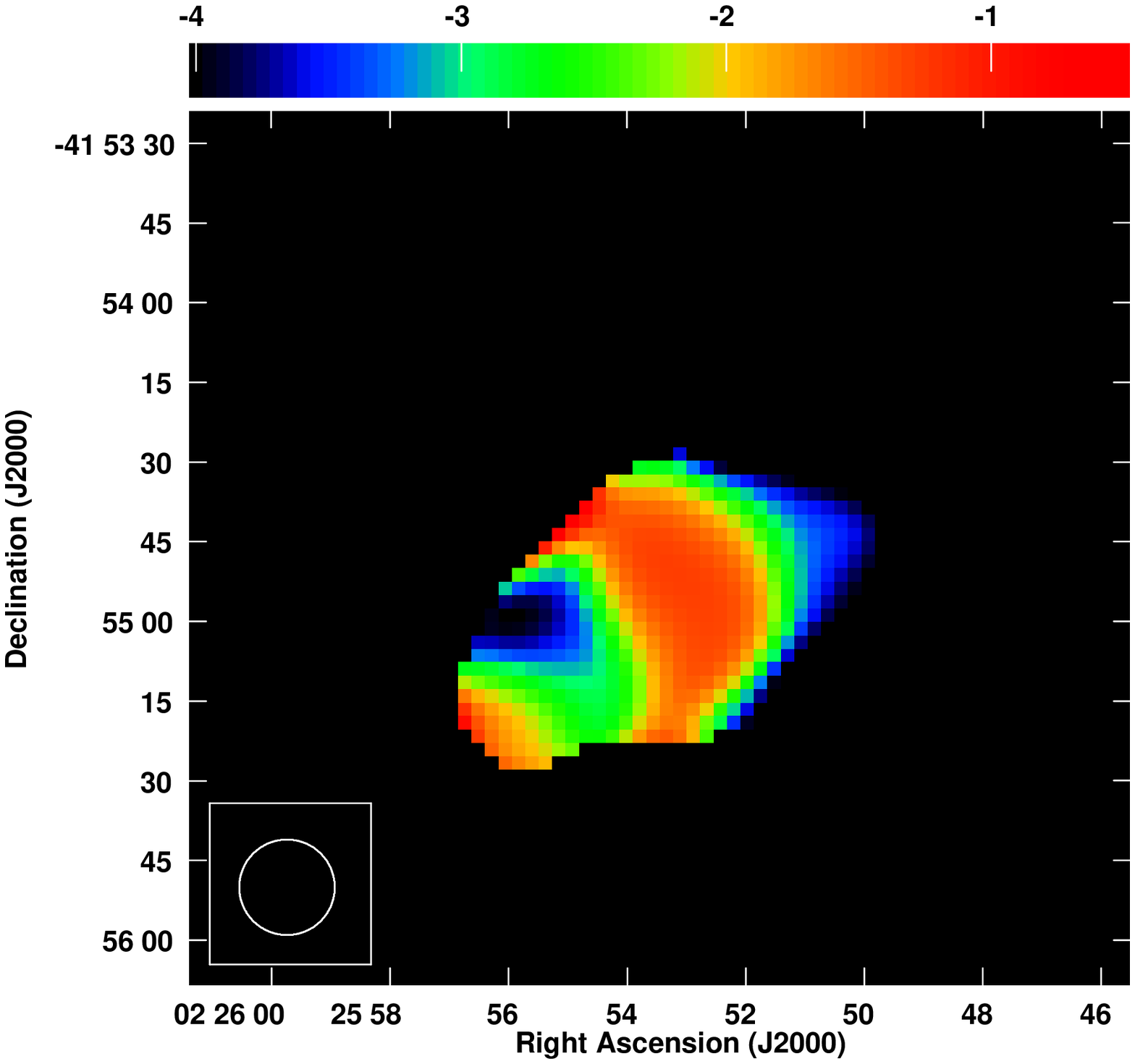}
   \caption{Left panel: In-band spectral index map of Abell 3017, created using uGMRT Band-5 data; Right panel: uGMRT 384-349 MHz spectral index map of Abell 3017. }
\label{fig:spix}
\end{figure*}

\subsection{X-ray spectral analysis}

\subsubsection{X-ray Cavities and Excess Emission}
\label{STE}
We have performed  spectral analyses of three specific regions. X-Ray Spectra were extracted from the two X-ray cavities,  marked as white ellipses in (Fig.~\ref{fig5}), and from the region of diffuse extended excess emission towards the South of the core marked ``EE" (panda region highlighted in blue). The extracted spectra are fitted with absorbed single temperature {\tt (phabs*apec)} models, with the redshift of the sources fixed at the redshift of the cluster ($z\!=\! 0.219$). The best-fit parameters are tabulated in Table~\ref{tab4}. This leads to an X-ray luminosity for ``EE" is found to be $(8.91\pm0.05)\times10^{43} {~\rm erg~s^{-1}}$.

\subsubsection{Deprojected X-ray spectral analysis}
In order to find the cooling radius, r$_{\rm cool}$, the bolometric luminosity confined within the cooling radius, L$_{\rm cool}$ and the corresponding cooling time, t$_{\rm cool}$, we deprojected the spectral information to estimate the parameters in 3D.  The cooling radius is estimated from the cooling time profile. It is defined  as the radius within which the cooling time equals 7.7 Gyr~\citep{2004ApJ...607..800B}. The bolometric luminosity within the cooling radius is equal to the cooling luminosity. We extract the spectra from seven concentric annuli by keeping S/N ratio at 35. We convolved the {\tt (projct)} model with  {\tt (phabs*apec)} model to obtain simultaneous best fitting model parameters satisfying spectral data from all annuli. From the deprojected temperature, abundance and density, we obtained the bolometric luminosity and cooling time for \mac~and the resultant  profile is shown  in Fig.~\ref{CT}, where the horizontal black line at 7.7~Gyr represents the cosmological time (or age of the cluster), and the corresponding radius  is called  the cooling radius  $r_{\rm c}$~(deproject) magenta color~ \citep{2004ApJ...607..800B}. The cooling radius  for the cluster \mac\  is found to be $\sim$102\,kpc (i.e. $t_{\rm cool}$ at 102\,kpc = 7.7~Gyr). We estimated the cooling luminosity within the cooling radius to be $L_{\rm cool}= (2.18\pm0.12)\times10^{44}$\lum.

\begin{table}
\caption{Properties of the spectra extracted from the regions discussed in \S.\ref{STE}. For each region we indicate its name, gas temperature and norm with \(1\)-\(\sigma\) confidence errors, the best-fitting reduced \(\chi^2\).}
\begin{center}
\begin{tabular}{crcccc}
\hline
\hline
Region      &      \(kT\)~~            &  Norm ($10^{-4}$)          & \(\chi^2\) (d.o.f.) \\
            &   \(\mbox{ (keV)}\)          &    ${\rm cm^{-5}}$                        &                  \\
\hline
Ecavity     &   \({4.45}_{-1.41}^{+1.29}\)    & \({0.92}_{-0.31}^{+0.71}\) &19.90  (15)   \\
Wcavity     &   \({3.59}_{-1.69}^{+1.32}\)    & \({0.92}_{-0.18}^{+0.19}\) &16.21  (13)   \\
EE          &   \({4.89}_{-0.31}^{+0.39}\)    & \({5.58}_{-0.31}^{+0.40}\)  &54.82  (64)   \\
\hline
\hline
\label{tab4}
\end{tabular}
\end{center}
\end{table}

\section{Radio images and spectra}
\subsection{Radio Morphology}
Our uGMRT images show three distinct features, namely the central BCG, the east and the west lobes and an unknown source beside the west lobe. While the emission from the lobes is not evident in the Band-5 image, it is seen in all the sub-band images of Band-3. The emission from the unknown source is prominent in all the sub-bands of both Band-3 (Fig.~\ref{fig:pband}) and Band-5 images (Fig.~\ref{fig:lband}). The unknown source has an arc-like morphology which is reminiscent of radio relics found in merging galaxy clusters. The source was previously identified by \cite{2017MNRAS.470.3742P} from earlier dual-frequency 610/235 MHz band GMRT observations (obsid. 28$\_$087), although they did not elaborate on the details of the source.

\subsection{Radio Spectral Index}
As described in \S~\ref{dataan}, we have generated a total of seven sub-band images from the uGMRT data at Band-3 and Band-5. The spectral index values for different components are given in Table~\ref{tab6}. Since the central source is a point source, the flux of the BCG is reliable despite the difference in resolution in both the bands and all the seven data points can be used to estimate the spectral index. We use the peak flux estimated from fitting a Gaussian model to the point source to avoid contamination from the lobes. The spectrum of the BCG is shown in Fig.~\ref{fig:spectral_index_bcg}.  The spectrum of the central source is linear, the spectral index being -1.4, indicating that \mac~is a steep-spectrum source. 

The lower resolution spectral index map made using Band-3 data (see Fig.~\ref{fig:spix}) shows a steepening trend along the lobes. The ultra-steep spectrum seen in the lobes is in good agreement with the past low-frequency observations by \cite{2017MNRAS.470.3742P}. However, the end of the south-west lobes, which coincides with the unknown source, has a flatter spectrum compared to the lobes. The emission from the lobes is not seen in the Band-5 images, likely due to the steepness of the spectrum. The unknown source possesses a spectral index comparable to that of the core. Both the Band-3 and Band-5 spectral index images show a similar average spectral index value of $-1.8$ for the unknown source.

\begin{table}
\centering
    \caption{Radio spectral fitting parameters}
    \label{tab6}
    \begin{tabular}{cc}
        \hline
        Region & $\alpha$ \\
        \hline    
        West Lobe & -3.4\\
        East Lobe & -3.5\\
        BCG & -1.4\\
        Unknown Source & -1.8\\
        \hline
    \end{tabular}
    \end{table}

\section{Discussion}
\begin{figure*}
\centering
\includegraphics[width=170mm,height=70mm]{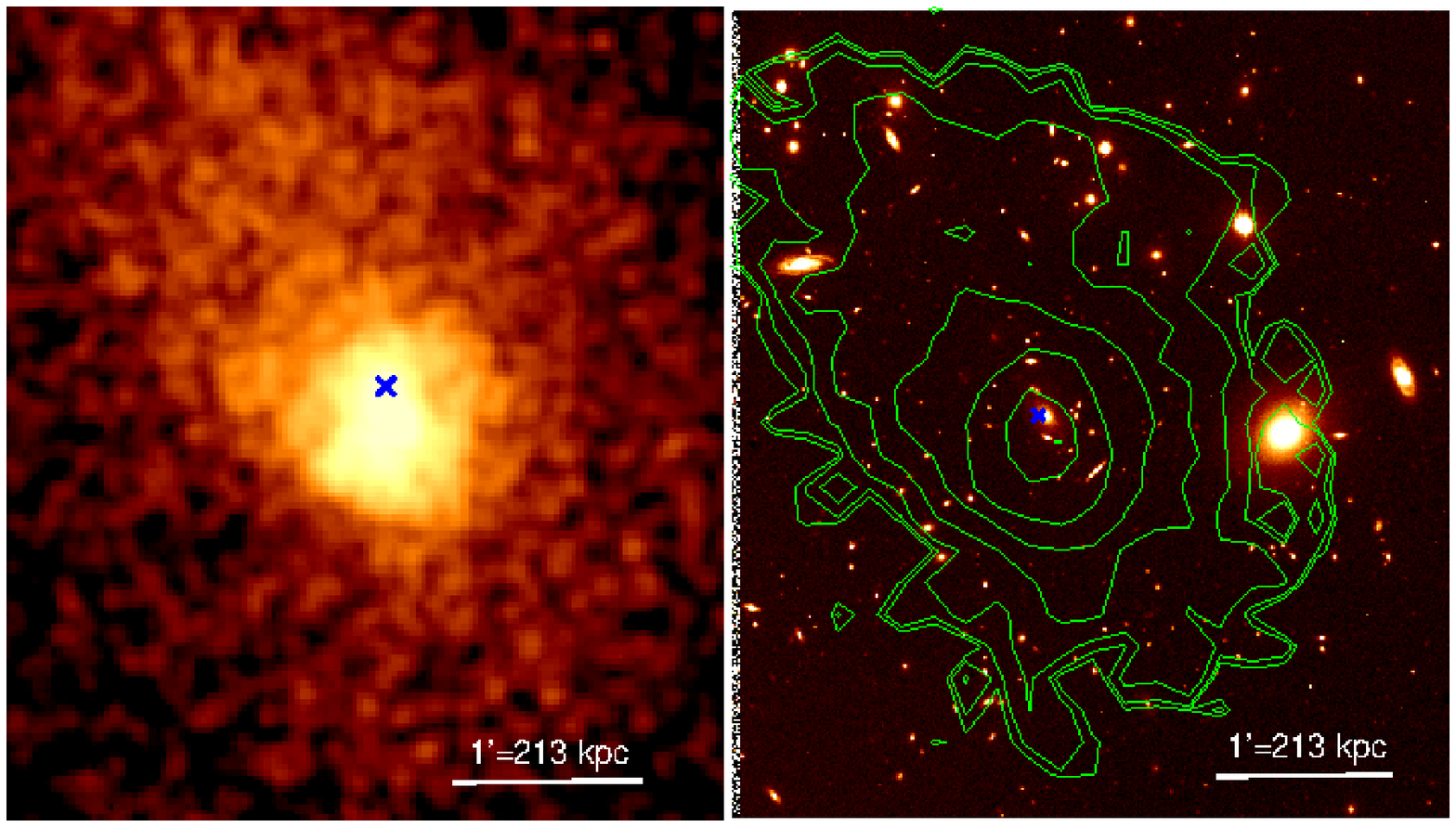}
\includegraphics[width=170mm,height=70mm]{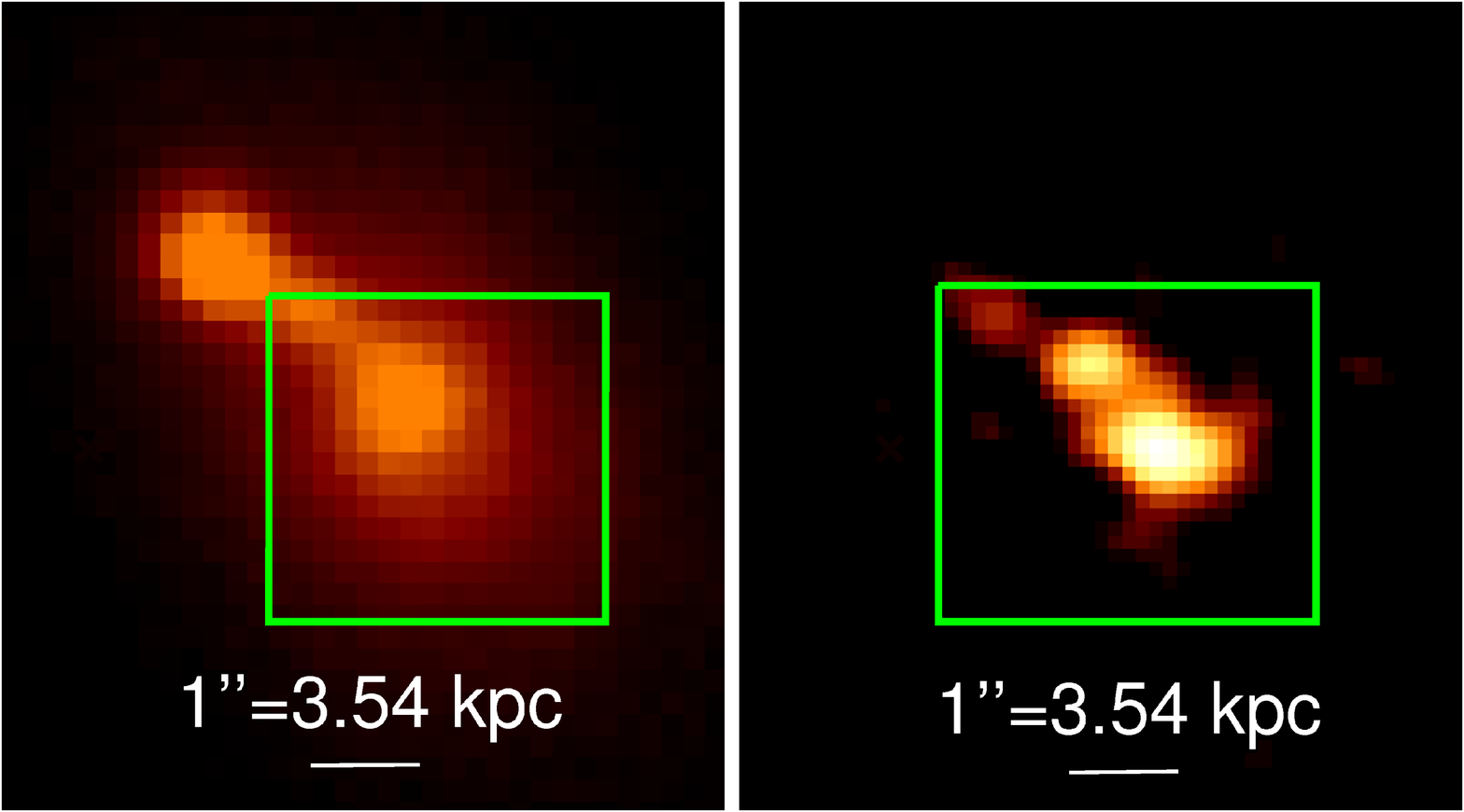}
\caption{{\it Top left panel:}  {\it Chandra} image of \mac~in the energy range of 0.5$-$4.0\,keV band (exposure-corrected, background subtracted, point sources removed), 4$\times$4 \arcmin, smoothed by a Gaussian width $\sigma$=3\arcsec. {\it Top right panel:} ESO/VLT R-band optical image,  with \chandra\ surface brightness contours overlaid.
{\it Bottom left panel:} Central 5\arcsec$\times$5\arcsec $R$-band optical image from  VLT/FORS1.  {\it Bottom right panel:} The near-infrared red-shifted  ${\rm~Pa\alpha}$  (2.2866 $\mu m$) emission map from the {\tt SINFONI} $K$-band  image. The 2.5 $\arcsec \times 2.5\arcsec$ green box shown on each image to compare near Infra-red and optical emission regions.}
\label{mor}
\end{figure*}

\subsection{X-ray and Optical Morphology}
\label{Sec: XOM}

In order to visualise the diffuse X-ray emission in the core of the cluster, and its relation to the 
 central sources and their optical counterparts, we use the central $2\arcmin \times 2$\arcmin~ X-ray image from \chandra\ and the optical $R$-filter image from the archival FORAS1 instrument on ESO-VLT, as shown in Fig.~\ref{mor}. The X-ray surface brightness contours from the \chandra\ 0.5$-$4.0\,keV image are overlaid on the optical image to compare the spatial extent of optical and X-ray photons. From the image (Fig.~\ref{mor}, top left panel)  it is evident that the central X-ray emitting gas is not uniformly distributed, but shows spiral-like feature towards the west of the cluster (for more details see \cite{2017MNRAS.470.3742P}, their Fig.~1b). This could be due to the sloshing of hot gas following the recent merger.
 
 We compared the central 5\arcsec $\times$ 5\arcsec
  optical image in the FORS1 R-band image with the same region in the near-IR, from the {\tt SINFONI} K-band ($1.9-2.4\micron$) emission. These images are shown in Fig.~\ref{mor} (bottom left and right panels). From these images, it is evident that the near-IR emission comes from the south-west component of the central source. To understand the detailed merger dynamics, optical spectroscopic observations, preferably using an IFU, are needed.

\subsection{The Central AGN}
It is now well understood that powerful radio jets emanating from supermassive black holes (hereafter SMBHs) play a vital role in the formation of X-ray cavities, excess X-ray emission, large-scale shock fronts and other substructures observed in the cluster environment \citep[e.g.][]{2005Natur.433...45M}. We can estimate the mass of the SMBH at the core of the  BCG of this cluster, using the well$-$established $M_{BH}-\sigma$ correlation between the black hole mass and the central stellar dispersion velocity of galaxies \citep[e.g.][]{2009ApJ...698..198G}. 
For \mac, we used the FWHM value measured from the central $\sim$ 2\arcsec $\times$ 2\arcsec  spectrum to calculate the 3 dimensional  stellar dispersion velocity from \cite{2016MNRAS.460.1758H}, and  it is found to be equal to 269$\pm$78 km s$^{-1}$. The modified $M_{BH}-\sigma$ correlation from \citep{2009ApJ...698..198G} leads to the central black hole mass of $\sim4.5 \times 10^8~{\rm M_{\odot}}$.  Accretion of matter on to the SMBHs is responsible for the AGN outburst in the form of radio jets and hence the formation of X-ray cavities \citep{2014MNRAS.442.3192V}.

We estimated mass accretion rate $\dot{M}_{\rm acc}$ on to the SMBH in BCG of \mac\ by using the relation suggested by \cite{2006ApJ...652..216R} and found it to be $ 0.12~\Msun yr^{-1}$.  Following \citep{2002apa..book.....F,2006ApJ...652..216R,2016MNRAS.461.1885V}, we estimated the ratio of  $\dot{M}_{\rm acc}/\dot{M}_{\rm Edd}$ for the central SMBH as  1.29 $\times\, 10^{-2}$, slightly equal or above  the threshold  limit set by \citep{1994ApJ...428L..13N} for radiatively inefficient accretion flow, 0.01~$\dot{M}_{\rm Edd}$.

\subsection{Excess Emission}
{We detect an excess of diffuse extended X-ray emission towards the South of the core of Abell~3017, which is  marked as ``EE" (panda region highlighted in blue, in Fig.~\ref{fig5}). A visual inspection of the background subtracted, exposure corrected 0.7$-$2.0 keV image shows some structure around the core of cluster (see Fig.~\ref{fig2}) \citep{2017MNRAS.470.3742P}. We could not find any corresponding optical galaxies in the VLT archival $R$-band images of the region, and so this is not the signature of a galaxy group. Instead, we speculate that this excess X-ray emission is likely due to bulk motions in the ICM, either due to gas sloshing or ram pressure stripping during a minor merger.

\subsection{WISE mid-IR colour and Star Formation}
\label{SFR}
The central dominant (cD) galaxy of \mac, 2MASX J02255309-4154523, has been detected as a strong mid-IR source in all the four bands of the Wide-field Infrared Survey Explorer (WISE), centred at 3.4 (W1), 4.6 (W2), 12 (W3) and 22$\micron$ (W4). The WISE mid-IR colours for 2MASX J02255309-4154523 are estimated to be [W1-W2]=0.36, [W2-W3]=2.45, [W3-W4]=2.25 and [W2-W4]=4.69. The bluer-than-usual colour index of [W1-W2]$\sim 0.36$, compared to the redder colour index of [W2-W3]$\sim$2.45 indicates that the cD galaxy of \mac\ has active star-formation \citep{2014ApJ...788..174B}. 

We looked for evidence of cool molecular gas in the core of the central galaxy, from the {\tt SINFONI} archival near-infrared  data ({\S \ref{IRI}}). The {\tt SINFONI} {\it K}-band  2.2866~$\mu m$ ${\rm Pa\alpha}$ image (Fig.~\ref{mor} bottom right panel), shows the presence of bright extended emission, within the central $\sim$3\arcsec $\times$ 3\arcsec.

 Such emission is from the inner few kiloparsec and could come from the warm (T $\sim$ 1000$-$ 1500 K) molecular material which is being deposited from the outer part of the cluster to the center of the cluster \citep{1998ApJ...494L.155F}. This ${\rm~Pa\alpha}$ could be related to the interaction between the radio jets and the warm  molecular gas emission which is being deposited in the form of cooling-flow. This confirms that the detection of molecular gas in the core of \mac\ related to a star formation rate.

We attempted to quantify the star-formation rate by using scaling relations from \citealt{1998ARA&A..36..189K}, using the           ${\rm H\alpha}$ emission line luminosity \citep[][${L_{\rm H\alpha}}\sim (110\pm17) \times 10^{40}\,{\rm erg s^{-1}}$]{2016MNRAS.460.1758H}   and  the far-ultraviolet (FUV) luminosity (${L_{\rm FUV}} \sim (2.10\pm0.18) \times 10^{44}\,{\rm erg s^{-1}}$) from {\it GALEX}. This yielded values of star-formation rate of $\sim 5.1\pm0.8$ \Msun yr$^{-1}$ and  $\sim 9.2\pm0.8$ \Msun yr$^{-1}$, estimated from the ${\rm H\alpha}$ and FUV luminosities, respectively.

\begin{figure}
\includegraphics[width=90mm,height=90mm]{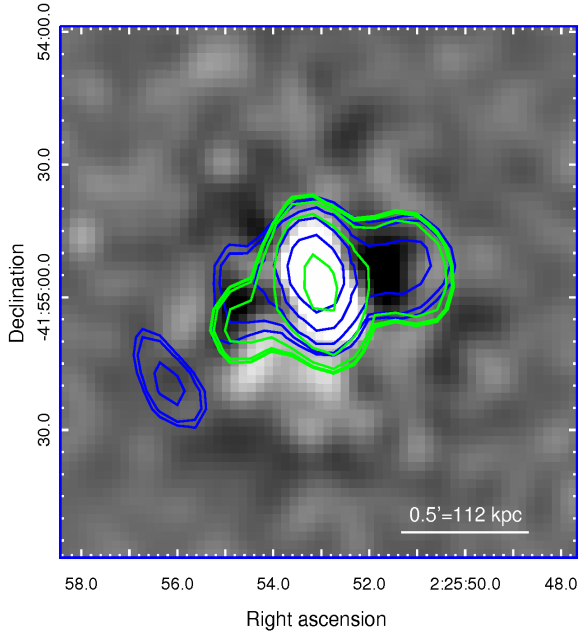}
\caption{{\it Chandra} X-ray central core region of \mac, in the energy range of 0.5$-$7.0\,keV, on which GMRT 235~MHz (green) and GMRT 610~MHz (blue) contours are overlaid. The lowest radio contour is drawn at 5$\sigma$ above background, with subsequent level increases in multiples of $\sqrt{2}$. At 235~MHz, 1$\sigma$ the rms is $\sim$ 0.45 mJy/beam, while at 610~MHz, the 1$\sigma$ rms is $\sim$ 60 $\mu$Jy/beam. Both radio images are at original resolution.}
\label{CR}
\end{figure}  

\begin{figure*}
\includegraphics[width=85mm,height=85mm]{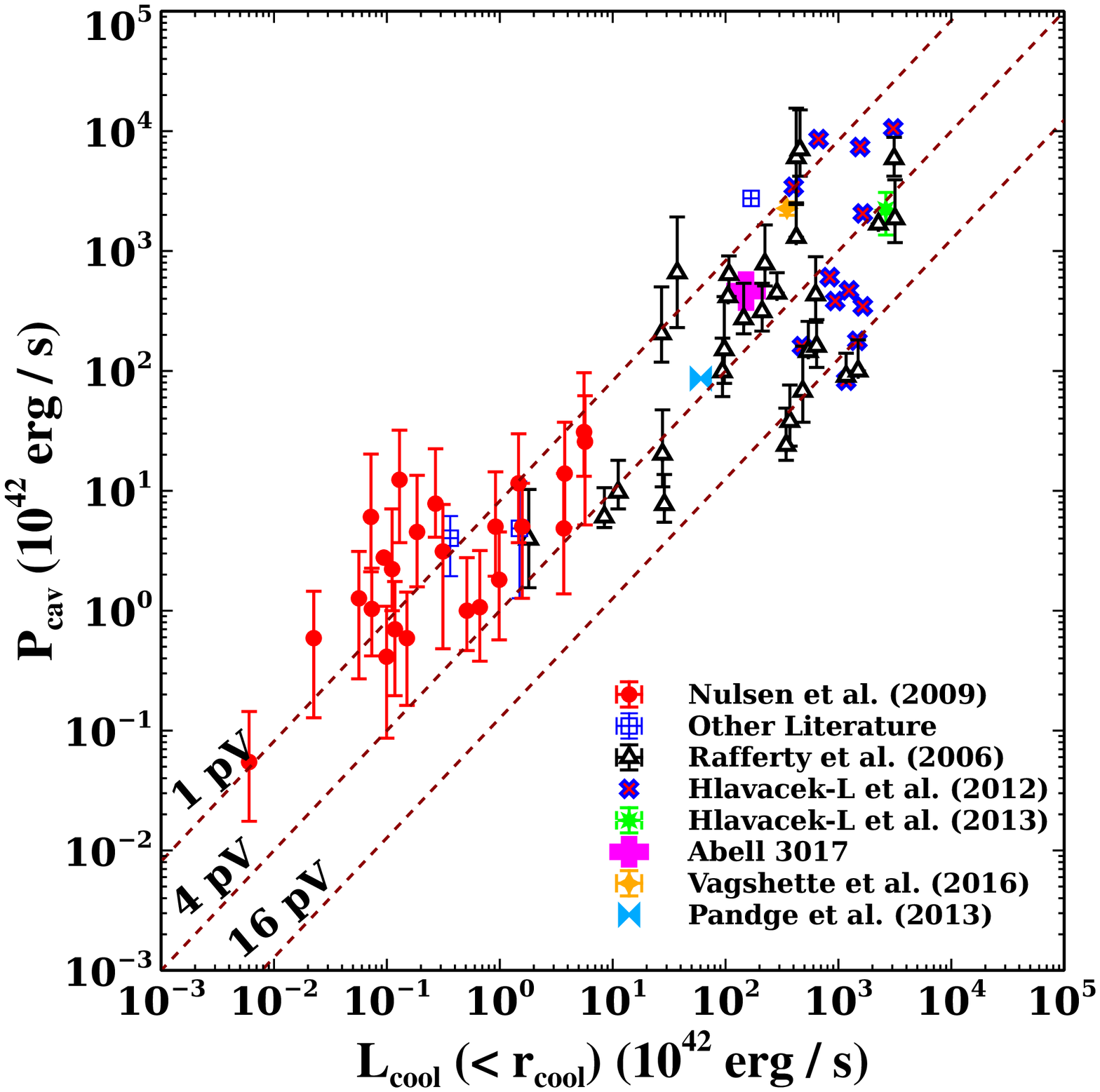}
\includegraphics[width=85mm,height=85mm]{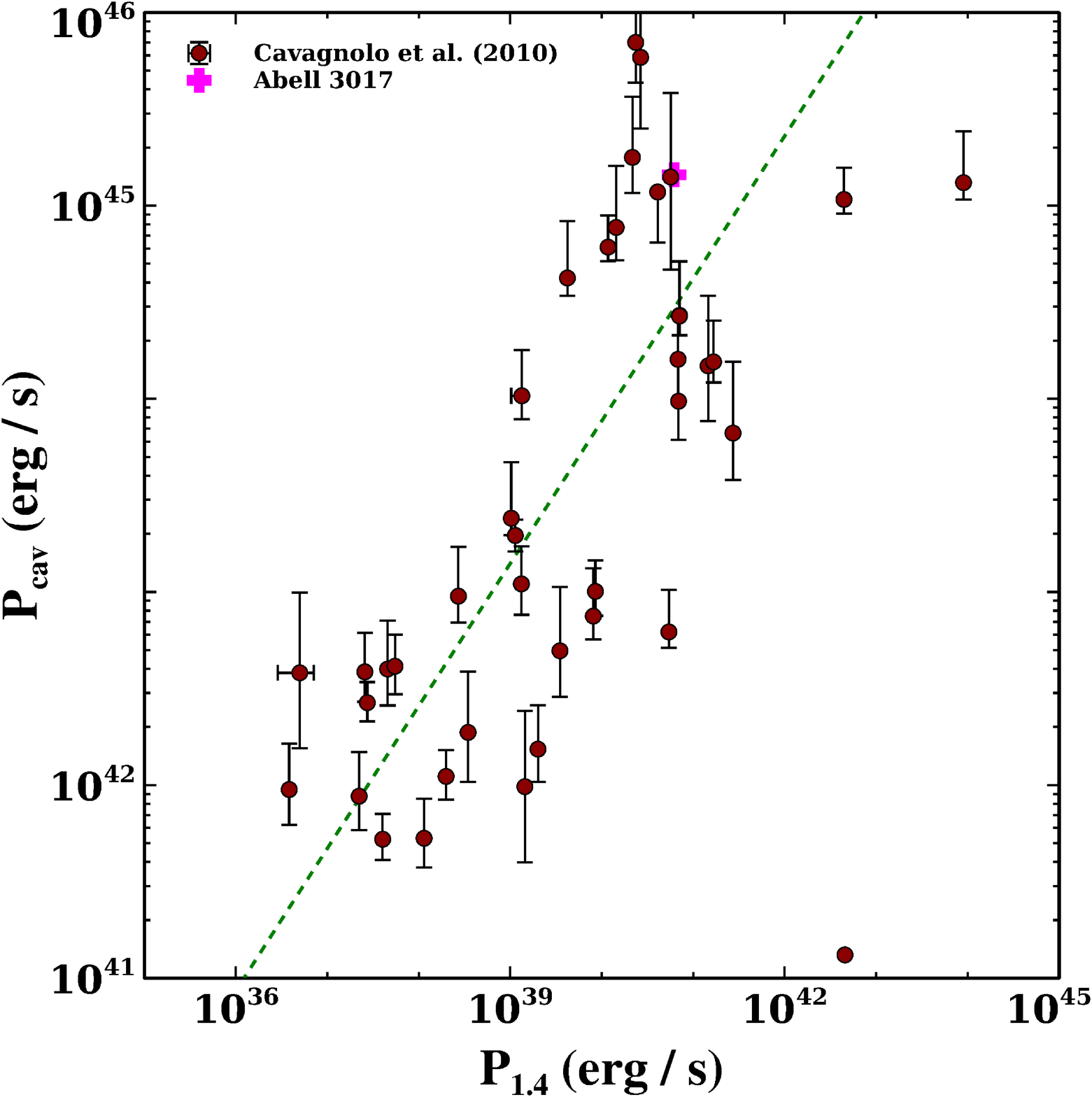}
\caption{{\it Left panel:} Cavity power versus the X-ray luminosity within the  cooling radius [points from  \protect\cite{2006ApJ...652..216R} are shown as open triangles (black), while those from \protect\cite{2009AIPC.1201..198N} are represented by filled circles (red)].  The cavity power and X-ray luminosity for \mac~is indicated by a plus sign (magenta).  The diagonal lines represent samples where $P_{\rm cav}=L_{\rm cool}$, assuming $pV$, $4pV$ and $16pV$ as the total enthalpy of the cavities. {\it Right panel:} Cavity power vs. radio power, where filled circles represent the galaxy clusters and groups from \protect\cite{2010ApJ...720.1066C}. The magenta plus symbol represents the \mac~cluster (from this work).} 
\label{fig9}
\end{figure*}
\begin{table*}
\caption[]{X-ray cavity properties of ~\mac}
\resizebox{12cm}{!} {
\begin{tabular}{lcccccccccc}
\hline
\hline
Cavity   & $a$   & $b$   & $R$   & $t_{\rm cs}$ &   Vol                 & $P_{\rm cav}$            & $E_{\rm Cavity}$  \\
         & (kpc) & (kpc) & (kpc) & ($10^7$ yr)  & $10^{69}~{\rm~cm^{3}}$ &($10^{44}$ erg $s^{-1}$) & ($10^{59}$ erg)  \\
\hline
W Cavity & 39.28 & 29.8  & 57    & 3.3          &   1.09    & \({5.69}_{-1.81}^{+2.92}\)           & 5.93  \\
         &       &       &       &              &           &                                      &         \\
E Cavity & 42.22 & 31.4  & 71    & 3.9          &   1.28    & \({5.53}_{-1.90}^{+2.86}\)           & 6.46  \\
\hline
\end{tabular}}
\label{cavtab}
\end{table*}

\subsection{Interaction of X-ray and Radio plasma}
We also studied the association of extended radio emission from the core of \mac\ with the detected X-ray cavities. To find X-ray and radio emission distribution in the cavity regions, we used the unsharp-masked X-ray image from current analysis and the radio emission images from \cite{2017MNRAS.470.3742P} as shown in Fig.~\ref{CR}. From this figure, it is evident that GMRT~235 and~610~MHz radio emission fills the X-ray cavity regions. Similar X-ray cavities have been reported in the cores of several groups and clusters of galaxies and are believed to be a signature of interaction of the AGN jets with the surrounding ICM. \citep{2009ApJ...693...43G,2009ApJ...705..624D,2010ApJ...714..758G,2012MNRAS.421..808P,2015Ap&SS.359...61S,2015ApJ...812..153O,2016MNRAS.461.1885V,2020MNRAS.tmp.1675K}.

\subsection{Cavity Energetics}
\label{CE}
In the case of \mac, we also compare the energy injected into the hot gas by the central AGN, to the energy lost by the hot IGM due to X-ray radiative process. The power contained within the X-ray cavities can be estimated from the measurement of the work done ($pV$) by the radio jet, in inflating the cavity, together with an estimate of the cavity age \citep{2007ARA&A..45..117M,2010ApJ...714..758G}, i.e. by measuring cavity enthalpy $4pV$ (for relativistic plasma) and age of the cavity. Total cavity enthalpy can be estimated using the relation \citep{2004ApJ...607..800B,2006ApJ...652..216R}, $E_{\rm cavity}=\frac{\gamma }{\gamma-1}$ pV;
where {\it V} represents the volume to the cavity, and given by V=4$\pi a^2$b/3, a and b are the semi-major and semi-minor axes
and {\it p} is the pressure exerted by the radio jets on the surrounding ICM.  

The gas pressure ({\it p=nKT}, where n=1.92 $\it n_{\rm e}$) surrounding to the radio jets/cavities was obtained from the deprojected temperature and density values  of the plasma.

The deprojected pressure was estimated by fitting the universal galaxy cluster pressure profile \citep{2010A&A...517A..92A} to the deprojected data points. The best fitting universal pressure profile was then used to interpolate the pressure value at the outer end of each of the cavities. This pressure value is used to estimate cavity power associated with each cavity on either side of the cluster. 


In order to estimate the age of each cavity, we have followed the procedure outlined in \cite{2004ApJ...607..800B}. The cavity age is estimated as the sound crossing time $t_{\rm c_s}$ (i.e. the time required for the cavity to rise the projected distance from the centre of AGN to its observed location with speed of sound) is, \[ t_{\rm c_s}\, = \frac{R}{c_{\rm s}}\, = R\,\sqrt{\mu\,m_H/\gamma\,kT},\],  where $R$ is the projected distance from the centre of cavity to the centre of AGN, $\gamma =4/3$, and $\mu = 0.62$. The  power of the cavity is estimated  using  \[ P_{\rm cav} = E_{\rm cavity}\, /  t_{\rm c_s}, \] where $E_{\rm cavity}$ is the cavity enthalpy and $t_{\rm c_s}$ is the age of the cavity. 

Uncertainties on the cavity power mainly arises due to projection of cavity along the line of sight culminating in an overestimated pressure value and an over or under estimation on the cavity volume. Table~\ref{cavtab} lists the volume, pressure, timescale, the resultant cavity powers along with the uncertainties in calculations for the two cavities.

Comparison of the total cavity power (${P_{\rm cav}}= ({11.22}_{-1.92}^{+2.40})\times10^{44}$\lum) and the cooling luminosity within the cooling radius  ($L_{\rm cool}= 2.18\pm 0.12\times10^{44}$\lum) values implies that AGN feedback through the radio jets is abundant to offset the radiative cooling within the cooling radius of this cluster.

We also derived the  X-ray cooling rate, within the cooling radius $r_{\rm cool}$, as   $\sim$ 200 M$_{\odot}$ yr$^{-1}$ ($\dot{M}=2L\mu m_{\rm p}/5kT$ \citealt{2012Natur.488..349M}). In Abell~3017, the radio jets originating from the AGN feedback results in the formation of prominent X-ray cavities in the ICM (see Fig.~\ref{CR}). As the jets inflate, the cool, low entropy, metal-rich gas in the ICM is pushed outward and forming the plume-like features  towards the southwest direction (see Fig. \ref{fig5}) similar to 3C320 \citep{2019arXiv190204778V}. It is known that such features around the  X-ray cavities can trigger star formation in the ICM. Therefore, we estimate the star formation rate (SFR) by using the ${\rm H\alpha}$ emission line luminosity  (${L_{\rm H\alpha}} \sim (110\pm17) \times 10^{40}\,{\rm erg~s^{-1}}$) and  ultra-violet (FUV) luminosity (${L_{\rm FUV}} \sim (2.10\pm 0.18) \times 10^{44}\,{\rm erg~s^{-1}}$) from {\it GALEX} (for more details see \S~\ref{SFR}). \cite{2015ApJ...811...73L} estimate the SFR over a wide range of cool core clusters, and obtain values from 0 to a few 100 M$_{\odot}$ yr$^{-1}$), with an average of 40 M$_{\odot}$ yr$^{-1}$.  The measured SFR in this cluster ($\sim 9.20\pm0.81$ \Msun yr$^{-1}$) is in agreement with \cite{2015ApJ...811...73L}, and is found to be approximately 4.6$\%$ of the cluster cooling rate. 

\subsection{Heating Versus Cooling of ICM}
In order to check that the AGN feedback evident in this cluster is efficient enough to quench the cooling flow, we compare the  derived total cavity power ($P_{\rm cav}$) with that of the total radiative luminosity derived within the cooling radius ($L_{\rm cool}$), as suggested by \citet{2006ApJ...652..216R} and shown in Fig.~\ref{fig9} (left panel). Solid diagonal lines in this plot represent the equivalence between the two ($P_{\rm cav}$ = $L_{\rm cool}$) at $pV$, 4$pV$ and 16$pV$ of the total enthalpy. Filled circles (red) represent the data points from the sample of \cite{2009AIPC.1201..198N}, while open triangles (black) represent sample data points from \cite{2006ApJ...652..216R}. The position of the \mac~is shown by the `$+$' symbol (magenta): it lies a little above the 4$pV$ line. 

\subsection{Power estimation from the Spectral Index}
The radio power was estimated using  \citep{kolokythas2018}
\begin{equation}
\label{eqn:1}
P_{\nu}= 4\pi D_{L}^{2} (1 + z)^{(\alpha -1)} S_{\nu}.
\end{equation}
The spectral index $\alpha$ is defined according to the convention $S_{\rm \nu} \propto \nu^{-\alpha}$ where $S_{\rm \nu}$ is the flux density at frequency $\nu$.  $D_{\rm L}$ is the  luminosity distance of the cluster. Further, we also compared the 1.4\,GHz radio power with the total cavity power $P_{\rm cav}$. For this, we plot the measured uGMRT 1.4\,GHz total radio power  ($\sim$ 7.94 $\times$ 10$^{40}$ \lum), 
with the total mechanical power of the cavities, as shown in Fig.~\ref{fig9} (right panel). In this plot, the long dashed line represents the best-fit relation of \cite{2010ApJ...720.1066C}, for a sample of giant ellipticals (gEs), with the upper and lower limits (1$\sigma$ confidence) indicated by the short dashed lines. In this plot, \mac~lies above the best-fit line obtained by \cite{2010ApJ...720.1066C}, suggesting that the radio source is capable of quenching the cooling flow. 

From the radio spectral images, we notice that a steep spectrum radio feature is visible $\sim$150\,kpc away from the central radio core of the \mac. We searched for an optical counterpart for the obvious feature, but cannot confirm any possible optical source in its vicinity. The imaging and spectral analysis further showed the shape of the unknown source to appear similar to a sharp arc. The flatter spectral index value of the unknown source, compared to that of the lobes, strongly indicates that this feature might be an old AGN pheonix  which had radiated away its energy, and has since been revived by the energy available from the shocked wavefront emerging as a result of the cluster merger \citep{2001A&A...366...26E,2004rcfg.proc..335K,2011MNRAS.414.1175O,2012A&A...546A.124V,2019ApJ...870...62P}. Such features are expected to be  $\leq 100$ kpc in size \citep{2004rcfg.proc..335K,2019ApJ...870...62P}, which is also true in this case. \citep{2017A&A...601A.145F,2019A&A...621A..77C} have also studied the \mac~cluster, using X-ray and optical data, and have shown the presence, in the ICM, of disturbances due to accretion shocks present in both clusters (Abell~3017 \& Abell~3016), as well as in the bridge region in between them. Their study supports our scenario that the radio phoenix detected in this system might have revived due to accretion shocks and/or turbulence present in the  ICM of this cluster.

\section{Conclusions}
We have presented a joint analysis of a combination of X-ray {\it (Chandra)}, optical (VLT ESO FORS1) and radio (GMRT \& uGMRT) observations of a merging galaxy cluster \mac. The objectives of the study were to identify and confirm possible X-ray cavities in the ICM of \mac, and to examine the energetics of AGN feedback and raditive cooling in the system. We summarise below some of the important results derived from this analysis:

\begin{enumerate}
\item 
The raw, unsharp-masked, contour-binned as well as 2-d $\beta$ model-subtracted residual images reveal  X-ray cavities (Ecavity and Wcavity), and an excess of X-ray emission around the centre of \mac. Surface brightness profiles derived for this cluster also confirm this. The X-ray cavities are located at projected distances of about 16\arcsec $\sim$(57\,kpc) and 20\arcsec $\sim$(70\,kpc) from the centre of \mac. \\

\item An excess of X-ray emission located at 25\arcsec $\sim$(88\,kpc) south of the centre of \mac, is likely due  to the bulk motions in the ICM either by gas sloshing or ram-pressure striping due to a merger.\\

 
\item The total mechanical power of the detected X-ray cavities $L_{\rm Cavity}$, and the X-ray luminosity within the cooling radius $L_{\rm cool}$, indicate that the total mechanical power emitted by central AGN is sufficient to balance the cooling loss in this cluster.\\

\item From the analysis of X-ray cavities and the cooling times, we estimate the mechanical power for both cavities to be equal to 
${P_{\rm cav}}= {11.22}_{-1.92}^{+2.40}\times10^{44}$\lum and $\sim$$L_{\rm cool}= 2.18\times10^{44}$~\lum, respectively. Comparing these values, we find that AGN feedback through the radio jets is sufficient to offset the cooling losses within the cooling radius of the cluster.\\

\item The current star formation rate of central BCG, from the ${\rm~H\alpha}$  and {\it GALEX} FUV luminosities, is estimated to be $\sim 5.06\pm0.78$ \Msun yr$^{-1}$ and $\sim 9.20\pm0.81$ \Msun yr$^{-1}$ respectively. \\

\item We detect a radio phoenix $\sim$150\,kpc away from the radio core, with a spectral index ($\alpha \!\leq\!-1.8$).\\

\item We also report the detection of the $\rm~Pa\alpha$ emission in this cluster, using ESO VLT {\tt SINFONI} imaging data.\\
\end{enumerate}

\section*{Acknowledgements}
The authors are grateful to the anonymous referee for their encouraging and constructive comments on the manuscript, that greatly helped us to improve its quality. MBP gratefully acknowledges the support from following funding schemes:  Department of Science and Technology (DST), New Delhi under the SERB Young Scientist Scheme (sanctioned No: SERB/YSS/2015/000534), Department of Science and Technology (DST), New Delhi under the SERB Research Scientist Scheme (sanctioned No: SERB/SRS/2020-21/38/PS). This research has made use of the data from {\it Chandra} Archive.   Part of the reported results is based on observations made with the NASA/ESA Hubble Space Telescope, obtained from the Data Archive at the Space Telescope Science Institute, which is operated by the Association of Universities for Research in Astronomy,  Inc., under  NASA  contract  NAS 5-26555. This research has made use of software provided by the Chandra X-ray Center (CXC) in the application packages CIAO, ChIPS, and Sherpa.  This research has made use of  NASA's  Astrophysics Data  System, and of the  NASA/IPAC  Extragalactic Database  (NED) which is operated by the Jet  Propulsion Laboratory, California Institute of Technology, under contract with the National Aeronautics and Space Administration.  Facilities: Chandra (ACIS), HST (ACS).
\section*{DATA AVAILABILITY}
The data underlying this article are available in the article and in its online supplementary material.

\def\aj{AJ}%
\def\actaa{Acta Astron.}%
\def\araa{ARA\&A}%
\def\apj{ApJ}%
\def\apjl{ApJ}%
\def\apjs{ApJS}%
\def\ao{Appl.~Opt.}%
\def\apss{Ap\&SS}
\def\aap{A\&A}%
\def\aapr{A\&A~Rev.}%
\def\aaps{A\&AS}%
\def\azh{AZh}%
\def\baas{BAAS}%
\def\bac{Bull. astr. Inst. Czechosl.}%
\def\caa{Chinese Astron. Astrophys.}%
\def\cjaa{Chinese J. Astron. Astrophys.}%
\def\icarus{Icarus}%
\def\jcap{J. Cosmology Astropart. Phys.}%
\def\jrasc{JRASC}%
\def\mnras{MNRAS}%
\def\memras{MmRAS}%
\def\na{New A}%
\def\nar{New A Rev.}%
\def\pasa{PASA}%
\def\pra{Phys.~Rev.~A}%
\def\prb{Phys.~Rev.~B}%
\def\prc{Phys.~Rev.~C}%
\def\prd{Phys.~Rev.~D}%
\def\pre{Phys.~Rev.~E}%
\def\prl{Phys.~Rev.~Lett.}%
\def\pasp{PASP}%
\def\pasj{PASJ}%
\def\qjras{QJRAS}%
\def\rmxaa{Rev. Mexicana Astron. Astrofis.}%
\def\skytel{S\&T}%
\def\solphys{Sol.~Phys.}%
\def\sovast{Soviet~Ast.}%
\def\ssr{Space~Sci.~Rev.}%
\def\zap{ZAp}%
\def\nat{Nature}%
\def\iaucirc{IAU~Circ.}%
\def\aplett{Astrophys.~Lett.}%
\def\apspr{Astrophys.~Space~Phys.~Res.}%
\def\bain{Bull.~Astron.~Inst.~Netherlands}%
\def\fcp{Fund.~Cosmic~Phys.}%
\def\gca{Geochim.~Cosmochim.~Acta}%
\def\grl{Geophys.~Res.~Lett.}%
\def\jcp{J.~Chem.~Phys.}%
\def\jgr{J.~Geophys.~Res.}%
\def\jqsrt{J.~Quant.~Spec.~Radiat.~Transf.}%
\def\memsai{Mem.~Soc.~Astron.~Italiana}%
\def\nphysa{Nucl.~Phys.~A}%
\def\physrep{Phys.~Rep.}%
\def\physscr{Phys.~Scr}%
\def\planss{Planet.~Space~Sci.}%
\def\procspie{Proc.~SPIE}%
\bibliographystyle{mn.bst}
\bibliography{mybib.bib}
\end{document}